\documentclass[preprint,3p,twocolumn,number]{elsarticle}

\usepackage{amssymb}
\usepackage{amsmath}
\usepackage{graphicx}
\usepackage{subfig}
\usepackage[colorlinks=true, allcolors=blue]{hyperref} 
\usepackage{comment}

\usepackage{lineno}

\journal{Powder Technology}

\begin{document}

\begin{frontmatter}



\title{Flow rate from a vertical silo with a tilted orifice}

\author[label1]{Ryan Kozlowski\corref{cor1}} 
\cortext[cor1]{rkozlows@holycross.edu}

\author[label2]{Luis A. Pugnaloni} 
\affiliation[label1]{organization={Physics Department, College of the Holy Cross},
            addressline={1 College Street}, 
            city={Worcester},
            postcode={01610}, 
            state={MA},
            country={United States}}
\affiliation[label2]{organization={Departamento de F\'isica, Facultad de Ciencias Exactas y Naturales,Universidad Nacional de La Pampa, CONICET},
            addressline={Uruguay 151},
            city={Santa Rosa, La Pampa},
            postcode={6300},
            country={Argentina}}

\begin{abstract}

The flow of dry granular materials from silos is of great practical interest in industry and of theoretical import for understanding multiphase dynamics. Recent studies have demonstrated that one way to control the rate of flow from a silo is to tilt it. However, this may not be practical in many industrial applications. Here, we demonstrate in experiments of quasi-2D silo discharge of monodisperse grains that the flow rate can be modulated by rotating the orifice – through elevating and shifting one side of the base – instead of tilting the entire silo. We use high-speed image analysis to track the average motion of grains in the silo. We first show that the flow rate decreases with orifice angle, but that this decrease is not as strong as when a silo is tilted or when a lateral orifice is used. However, with the addition of a grain-sized ridge on each side of the orifice, the flow rate collapses with prior tilted-silo results. We then characterize the flow velocity of grains exiting the orifice and highlight key features of the stagnant zones and slip zones on each orifice side. Finally, we model our results based on these measurements, demonstrating the importance of horizontal creep along slip zones next to the orifice and the narrowest opening cross-section through which the material flows. These findings reveal a simple method for controlling both flow rate and direction, and highlight the importance of both dynamics within and geometry of the stagnant zones near the orifice.

\end{abstract}

\begin{graphicalabstract}
\includegraphics[width=2.0\linewidth]{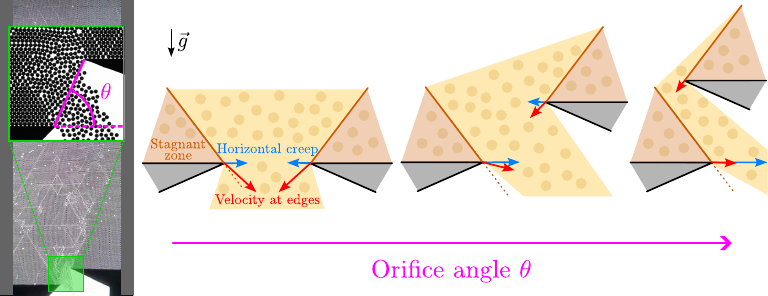}
\end{graphicalabstract}

\begin{highlights}
\item Vary orifice angle of quasi-2D silo by elevating one orifice edge.
\item Flow rate, average packing fraction, and average flow fields with high-speed imaging.
\item Effect of fixed grain placed on each orifice edge to minimize boundary slipping.
\item Horizontal creep at each orifice edge plays vital role in flow rate.
\item Stagnant zone constrains flow upstream from orifice for steep enough angles.
\end{highlights}


\end{frontmatter}


\section{Introduction}
\label{sec:introduction}

In many natural and industrial systems, particulate matter is driven through constricted openings, such as fine particles flowing through porous media~\cite{Liu2019}, gas hydrates flowing in pipelines~\cite{Sloan2011}, pedestrians exiting a room~\cite{Helbing2000,Helbing2005,Zuriguel2014b}, logs moving down a river~\cite{Gurnell2002}, or suspension particles navigating microfluidic channels~\cite{Haw2004,Genovese2011,Dressaire2017,Souzy2020}. Gravity-driven flows of granular materials~\cite{Jaeger1996} through a confined orifice in a silo are especially prevalent in industrial and geophysical systems, for example in the processing and transport of pharmaceuticals~\cite{Muzzio2002} and food products~\cite{Fitzpatrick2004,Job2009} and debris flows through harm-mitigating slit dams~\cite{Li2023}. Characterizing the flow rate and clogging behaviors of granular materials in silos is therefore important both for developing efficient practices~\cite{Muzzio2002,Saleh2018} and robust models of the rheology of intermittent dense granular flows. 

A large body of research confirms that the flow rate of dry, gravity-driven granular materials is described by the Beverloo Rule~\cite{Beverloo1961,Mankoc2007}, an empirical relationship for the flow rate of the material $Q$ based on dimensional analysis relating the relative size of grains $d$ to the orifice size $D$: 
\begin{equation}\label{eqn:beverloo}
    Q = C\rho_{\rm B} \sqrt{g} (D-kd)^{f/2} ,
\end{equation}
where $f = 3$ in two-dimensional (2D) silos and $f=5$ in three-dimensional (3D) silos, $C$ is a dimensionless parameter that is largely independent of details of the granular material~\cite{Darias2020}, $k$ is a parameter characterizing the so-called ``empty annulus'' accounting for steric limitations of grain flow near the orifice edges~\cite{Brown1970,Tighe2007,RubioLargo2015}, $\rho_{\rm B}$ is the bulk density of the granular material in the silo, and $g$ is the free-fall acceleration. For a flat-bottomed, upright silo, Beverloo's rule has successfully been modeled from first principles by considering energy balance in the silo~\cite{Madrid2018,Darias2020}. 

Experiments have shown that the flow rate of grains depends on a variety of factors related to the geometry of the container, including container dimensions~\cite{Lopez2020}, hopper angle~\cite{Rose1959,Darias2020hoppers,Mendez2021,Gago2023}, orifice shape~\cite{Zatloukal2012}, the presence of multiple orifices~\cite{Kunte2014}, and whether the orifice is on the bottom of the silo or on a side-wall~\cite{Chang1991,Medina2014,Zhou2017,Serrano2019,Anyam2022}. Ref.~\cite{Zhou2017} even found, surprisingly, that the Beverloo's rule power law exponent $5/2$ for a lateral orifice in a 3D silo becomes linear  ---not the expected $3/2$--- when the silo width is narrowed to approach a quasi-2D silo as the spacing confining the grains to a monolayer becomes a relevant lengthscale. 

The use of lateral orifices, or more generally orifices tilted away from the horizontal, is important in a variety of industrial settings. For example, one may need to modulate the direction of flow by tilting the opening or silo and then adjusting the flow rate with the orifice opening size. And in some cases, the spatial constraints of an existing industrial setup may actually prevent effective discharge from a bottom opening.

The first relationship between $Q$ and the inclination angle $\theta$ of the orifice above the horizontal was posed by Franklin and Johanson based on measurements for a few orientations: 
\begin{equation*}
    Q(\theta) = Q_0 \frac{\cos\theta_{\rm S} + \cos\theta}{\cos\theta_{\rm S} +1},
\end{equation*}
where the angle of the stagnant zone of the granular material is denoted by $\theta_{\rm S}$ and $Q_0$ is the flow rate at $\theta=0^{\circ}$~\cite{Franklin1955}. A dependence on $\cos\theta$ would suggest that the horizontal projection of the orifice determines the flow rate of the material, but since flows are observed to occur even for $\theta>90^{\circ}$, this is at best an incomplete, empirical model. Recent studies examining a more robust range of silo angles~\cite{Sheldon2010,Thomas2013} find that more elaborate models are necessary to capture $Q(\theta)$~\cite{Liu2014b}. In particular, Ref.~\cite{Kozlowski2023} examined a 2D tilted silo and modeled $Q(\theta)$ by considering the flow of side streams of the material along stagnant zone surfaces or, when tilted steeply enough, along one of the confining orifice walls. Despite their success in describing tilted silo discharge rates, the recently proposed models of Refs.~\cite{Liu2014b} and~\cite{Kozlowski2023} are not fully physically motivated: Ref.~\cite{Liu2014b} can only treat the regime $\theta>90^{\circ}$ semi-empirically, while Ref.~\cite{Kozlowski2023} proposes an empirical relationship for average outpouring flow speed with $\theta$. Moreover, the validity of existing model predictions has not been tested in other geometric configurations where the orifice is tilted without utilizing a side-wall or tilting an entire silo (where preparation protocol ---whether grains are poured prior to tilting, or if they are poured after tilting--- may be an important factor).

In this work, we present experiments studying the discharge of grains from a quasi-2D silo with varying orifice orientation by elevating one side of the bottom boundary vertically above the other while holding constant the edge-to-edge distance. This is a novel approach to tilting the orifice that may be more practically implemented than drilling new holes or tilting entire hoppers in industrial settings by using, for example, orifice inserts to change out the bottom boundary alone~\cite{Wiacek2023}. Since we find that in these configurations the typical stagnant zones are mobilized during discharge, we also run experiments adding a fixed grains to each edge of the orifice to prevent this flow contribution to some extent and help understand the dynamics. These scenarios allow for testing of prior models and, as we will motivate and show in this work, the development of a more robust, general, and physically motivated model. 

We first present the experimental methods. We next show results for the flow rate as a function of angle $Q(\theta)$ and, through particle tracking, the corresponding average velocity and packing fraction of the material in the vicinity of the orifice. In the process, we compare flow properties between our current results and those of past tilted silo studies, shedding insight into how orifice inserts may be used effectively to modulate flow. Finally, we develop a novel, fully physically motivated model that captures $Q(\theta)$.

\section{Experimental Methods} \label{sec:expmethods}

\begin{figure*}[h]
    \centering
    \includegraphics[width=0.90\linewidth]{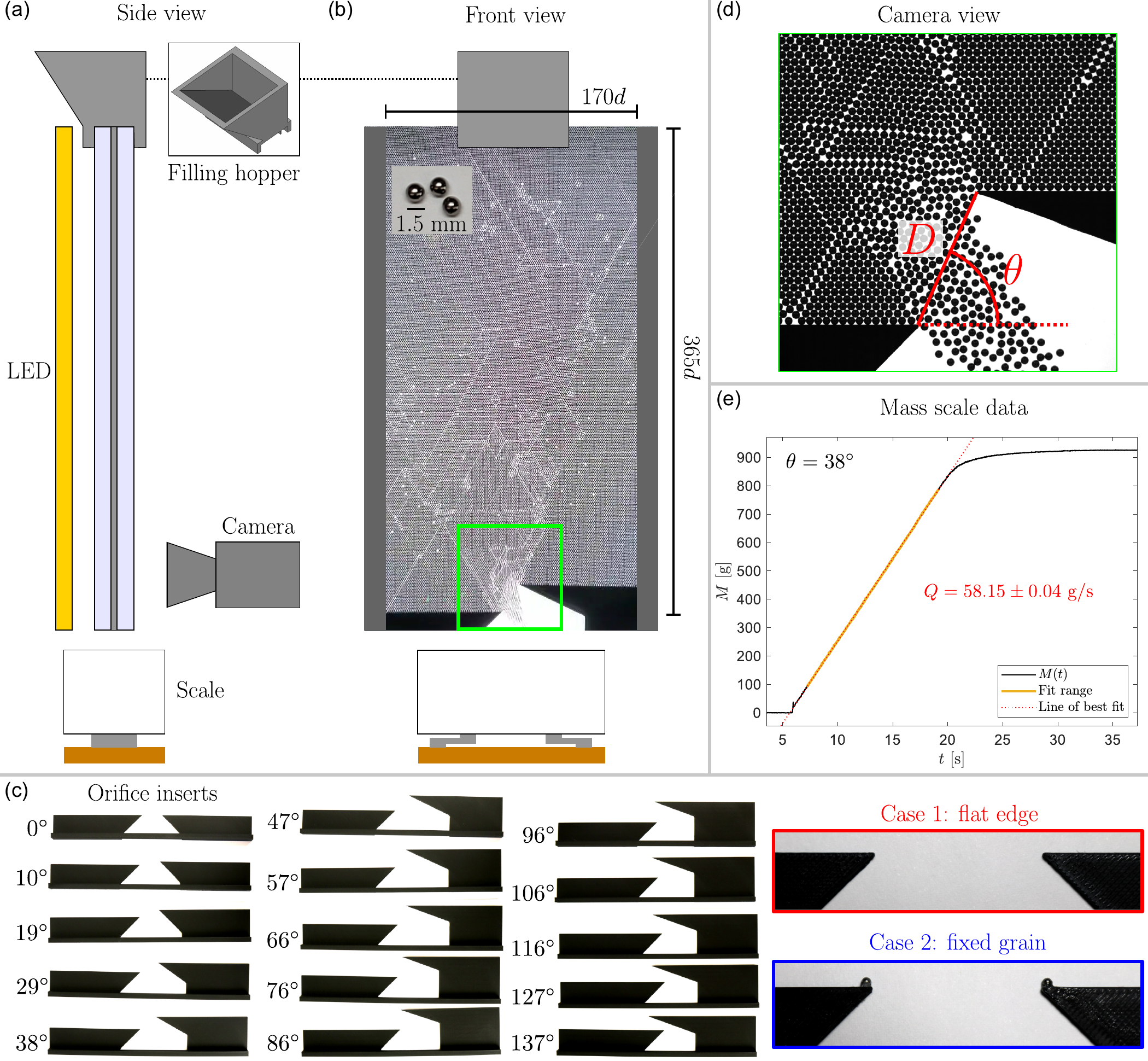}
    \caption{(a) Side view of silo. (b) Front view of silo. (c) Bottom boundary orifice inserts for all values of $\theta$ investigated. (d) Camera view of silo for $\theta = 66^{\circ}$ and $D=17.7d$. (e) Mass vs. time and line of best-fit for a fixed grain trial with $\theta = 38^{\circ}$.}
    \label{fig:schematic}
\end{figure*}

In our experiments, we use a quasi-2D silo with a monolayer of chrome steel ball-bearing grains (Gosensball 1.5mm G10 Hardened) between the front and back plates, as shown in Fig.~\ref{fig:schematic}a,b. We measured the diameter of 100 random grains in the sample with a manual micrometer of 0.001'' graduation (Miyutoyo 103-177) and obtained an average diameter $d = 1.4965\pm0.0019$~mm. The error corresponds to standard deviation of the measurements; the material is highly monodisperse with a size dispersity of 0.1\%. We measured the mass per grain $m = 0.0137\pm0.0001$~g with a scale of resolution 0.1~g (Ohaus YA501) by counting and measuring the total mass of several hundred beads, repeating this protocol three times with randomly sampled grains to obtain an average. The error in mass corresponds to the range across the three trials.

The silo is $170d$ wide and $365d$ high, and the orifice is centered at the bottom, characterized in this work by opening size $D$ and angle $\theta$ with respect to the horizontal. The silo is composed of two parallel, $0.5$-inch-thick static-dissipative acrylic plates (McMaster-Carr 8774K54) pressed together with aluminum spacers of thickness $1.59$~mm sandwiched in between the plates on the left and right sides. These spacers were chosen such that the average spacing between the plates is approximately $6\%$ larger than the grain diameter (with variations across the silo due to plate thickness variations) to prevent wedging of grains while also maintaining a monolayer. The top of the silo is open, and a custom 3D printed hopper is used to pour grains into the silo. 

To change geometric features of the bottom boundary, we 3D print different bottom boundary inserts using PLA with a Creality K1 Max 3D filament printer and sand the grain-contacting surface with 1000 grit sandpaper. As shown in Fig.~\ref{fig:schematic}c, the pieces are designed with one edge of the bottom boundary elevated above the other at fixed edge-to-edge distance $D = 17.7d$ to effectively rotate the orifice to angles $\theta\in[0^{\circ}, 140^{\circ}]$ without tilting the entire silo nor changing the orifice size. A connecting bar that extends out of the plane of the silo holds the two sides together without interfering with grains as they exit the silo. Each orifice side has an angled bottom side to prevent exiting grains from interacting with the boundaries below the orifice. The lower (leftmost in our images) orifice edge always has a $45^{\circ}$ angle on the bottom side, while the upper orifice edge (rightmost in our images) has a wedge angle that decreases from $45^{\circ}$ to a minimum $20^{\circ}$ at steeper tilt angles to avoid collisions of grains with the bottom side as they flow out. A wedge shape is necessary even for the upper boundary in order to have a stiff structure that does not deform under the weight of grains.

A custom-made mass scale below the orifice consists of a 3D-printed box on top of a wooden platform with two $780$~g load cells (CZL616) whose signals are amplified (HX711) and transferred to the operating computer via an Arduino serial port connection. The scale records a mass time series with a sampling rate of 80 Hz and resolution of $\pm0.5$~g, the mass of roughly 40 grains. Simultaneously, once mass data acquisition begins, a relay switch turns on an LED panel that back-lights the silo. After at least 1 second of flow, a button is pressed to trigger acquisition of images by a high frame rate camera (Photron Fastcam Nova-S6) with a macro lens (Sigma 105mm F2.8 EX DG OS HSM). The camera records the granular material in a $40d\times40d$ region in the vicinity of the orifice that is horizontally centered on the orifice, as shown in Fig.~\ref{fig:schematic}d. The recording rate is either 750, 1000, or 2000 frames per second as necessary for different trials in order to capture individual grain dynamics frame-by-frame (i.e., in order for the fastest grains to move less than a single grain radius frame-to-frame) while also maximizing the duration of recorded flow. Sample videos for each experimental condition are shown in the Supplemental Materials~\cite{SupplementalMaterials}. 

In a single trial, the appropriate orifice piece is fixed in the bottom of the silo and a cardboard plug is set in place. Grains are then poured into the silo with the hopper at the top, which is approximately 1/3 of the system width and held over the middle of the silo throughout the pouring process. The plug is then removed rapidly, grains flow from the orifice into the mass scale, the switch turns the LED on, and the camera is triggered. 

Five trials were conducted for each condition in this work; mass-time data was recorded in all cases, but high-speed video data was acquired for two out of five trials to reduce data volume.

\subsection{Two conditions: flat edge and fixed grain}

The orifice inserts we show in Fig.~\ref{fig:schematic}c have flat edges. We refer to this as the ``flat edge'' case. In order to evaluate the effect of the sliding of particles along the base, we implement a variation in the set-up that constrains this motion to some extent. We perform identical experiments using the same orifice inserts modified only by gluing a bead on each orifice edge, with the center of the grain positioned at a distance $d/2$ inward from the edge. We refer to this as the ``fixed grain'' case. The difference between these two conditions will be highlighted throughout this work.

\subsection{Mass-time data}

The mass as a function of time $M(t)$ acquired with the mass scale is used to obtain the average steady-state flow rate $Q$, as demonstrated by a sample time series in Fig.~\ref{fig:schematic}e. For each time series, we first determine the total range of mass data ($M_{\rm range} = M_{\rm final} - M_{\rm initial}$) registered by the scale. We then select a region of data to fit to a straight line for the time interval when $M(t)$ is between 10\% and 85\% of the total range. We have confirmed that our results are insensitive to the exact choice of the fitting range cutoffs as long as initial and final transients are avoided. Moreover, we have tested that the flow rate is in the steady-state regime for the entirety of the selected range by computing the slope in regular intervals within the range. Changes in the interval duration do not influence the results we present: any fluctuations in measured $Q$ due to parameter changes are within trial-to-trial fluctuations.

\subsection{Particle-tracking data}

With high-speed image data, we directly compute the frame-to-frame velocity of each grain in the vicinity of the orifice. In a single frame (frame 1), we identify particles to subpixel precision using standard morphological image operations such as erosion and grouping~\cite{Davies2018}. We do the same for the next frame (frame 2). We then construct a matrix where the $i$-th row and $j$-th column entry is the distance $d_{ij}$ between the particle $i$ in frame 1 and the particle $j$ in frame 2. We lastly solve the linear assignment problem with this distance matrix using Matlab's \verb+matchpairs+ function, in which the ``cost'' of distance is minimized via a matching algorithm~\cite{Duff2001}. We designate the cost of a particle \textit{not} being matched as $d/2$ to prevent particles from being matched if they are greater than one radius away from each other frame-to-frame.

\begin{figure}
    \centering
    \includegraphics[width=1.0\linewidth]{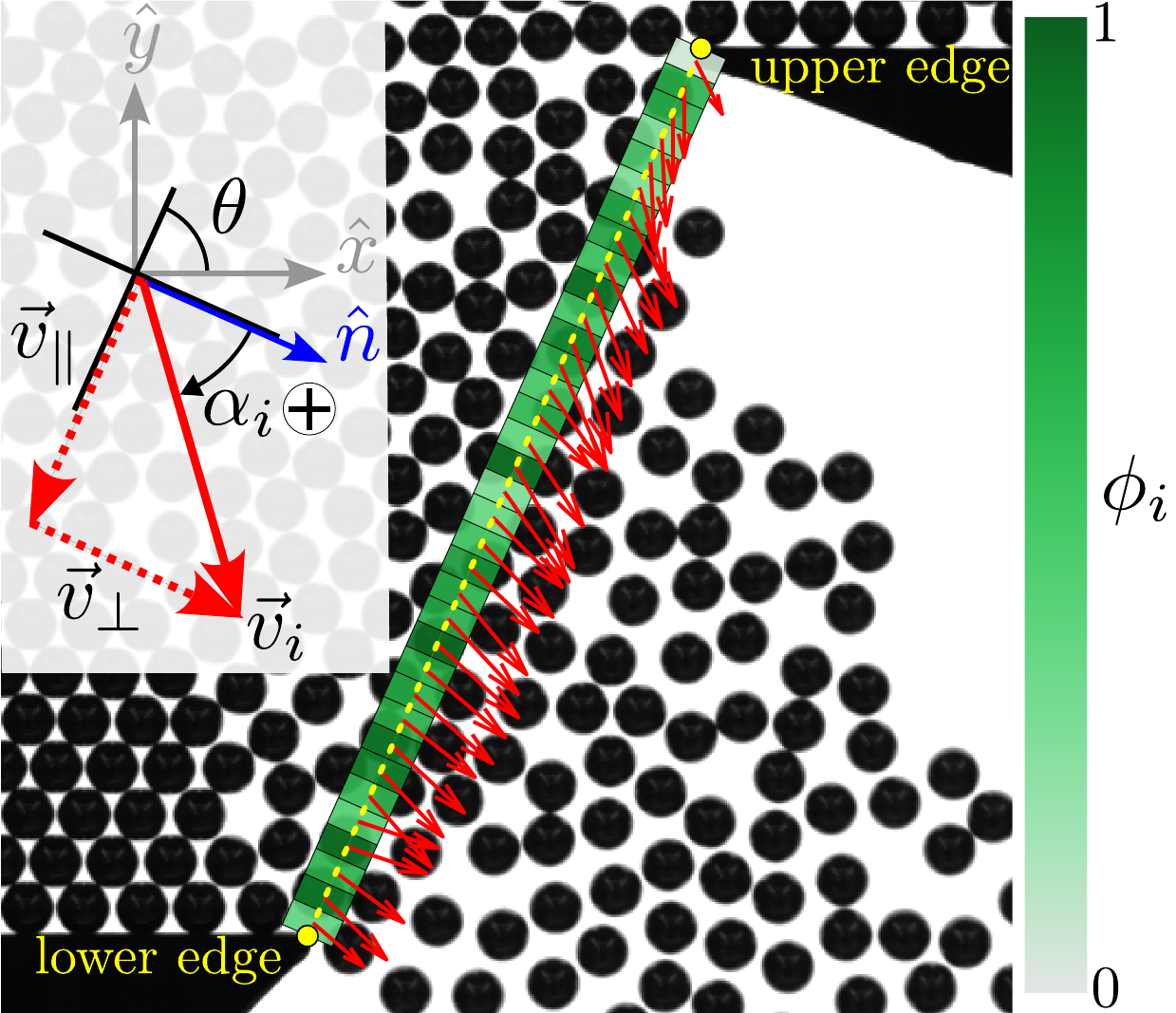}
    \caption{Example frame with velocity vectors $\vec{v}_i$ (red arrows) and packing fractions $\phi_i$ (color scale) for each bin $i$ along the orifice line. Inset: $\hat{n}$ is the orifice normal, $\theta$ is the magnitude of the angle between orifice line and $\hat{x}$, the outpouring velocity component is $\vec{v}_{\perp}$, the component of velocity along the orifice line is $\vec{v}_{||}$, and $\alpha$ is the angle between $\vec{v}$ and $\hat{n}$. In this example $\alpha$ is positive.}
    \label{fig:orificeprofile}
\end{figure}

With all particle position and velocity data in hand, we determine the average velocity and packing fraction profiles along the orifice. For a given trial, we first hand-select the orifice edges in one image. (We have confirmed, by randomly generating orifice edge-points in the vicinity of the selected point, that the results we present in this work are not influenced by the exact selection of the orifice edges.) We then construct 35 identical-sized sampling bins of width $0.506d\approx d/2$ spanning the length of the orifice and height $1d$ perpendicular to the orifice, as demonstrated in Fig.~\ref{fig:orificeprofile}. We sample every 10th frame from the original video, a time interval in which the fastest grains will move several grain diameters, in order to avoid any correlations between separate analyzed frames. For each analyzed frame, the local packing fraction $\phi$ in a bin is determined as the area of grains within the bin area divided by the bin area. The velocity $\vec{v}$ in a bin is determined as the weighted average of the velocities of any grains that intersect with the bin, where the weighting is the relative intersection area of each grain.  The average packing fraction and velocity in each bin is lastly calculated by an average over all analyzed frames.

Using this profile information, we calculate the flow rate 
\begin{equation}\label{eqn:qvideo}
Q_{\rm video} = \sum_{i}\rho \phi_i v_{\perp,i} \Delta x = \sum_{i}\rho \phi_i v_{i}\cos\alpha_i \Delta x
\end{equation}
where $\rho = m/A_{\rm grain} = 7.79$~kg/m$^2$ is the area density of a steel grain, $\phi_i$ and $v_{\perp,i}$ are the packing fraction and velocity component normal to the orifice of bin $i$, and $\Delta x \approx d/2$ is the bin width. We also extract the angle $\alpha_i$ at each bin, defined as the angle of $\vec{v}_i$ with respect to the orifice normal $\hat{n}$, as shown in Fig.~\ref{fig:orificeprofile}. Note that we will define $\vec{v}_{\perp}>0$ when along the same direction as $\hat{n}$, $\vec{v}_{||}>0$ when directed from the upper edge to the lower edge, and $\alpha>0$ when clockwise from $\hat{n}$. 

\section{Results}\label{sec:results}

\begin{figure}
    \centering
    \includegraphics[width=1\linewidth]{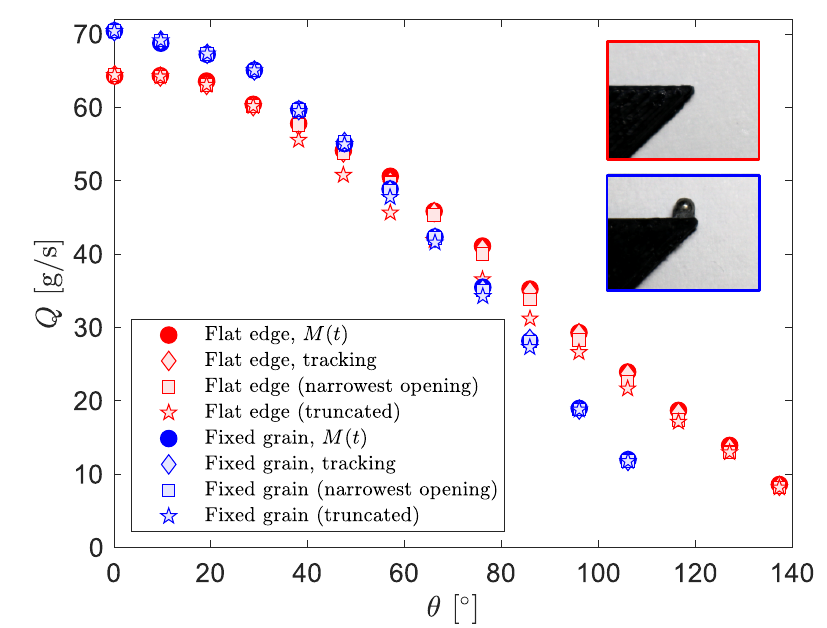}
    \caption{Average flow rate measured by mass scale data (filled circles) and image particle tracking analysis (open symbols) vs. orifice angle for flat edge and fixed grain cases. The tracking results were obtained with three different flow cross-sections: (diamonds) along the orifice line, (squares) along the narrowest opening cross-section extending across the image (see Sec. \ref{sec:modelnarrowest}), and (stars) along the narrowest opening but truncated at the stagnant zone boundary. Error bars represent the range across trials and are smaller than the data points. } 
    \label{fig:Q}
\end{figure}

Figure~\ref{fig:Q} displays the mass flow rate $Q$ for each experimental condition ---flat edge, and fixed grain--- as a function of orifice angle $\theta$. The data include the calculation of $Q$ by mass time series (e.g., Fig.~\ref{fig:schematic}e) (circles) from the scale and via particle tracking (diamonds) with Eqn.~\ref{eqn:qvideo}, which are in excellent agreement. (The other two cases, narrowest opening and truncated results, will be discussed in Sec.~\ref{sec:modelnarrowest}.) For the flat edge case, the flow rate decreases with angle until $\theta\approx 140^{\circ}$. Above this angle, the system becomes susceptible to clogging due to arches formed between the upper orifice edge and the stagnant zone~\cite{Garcimartin2010,Thomas2013,Zuriguel2014a}, which has an angle of approximately $60^{\circ}$ (see Sec.~\ref{sec:spatialflow}); we only examine cases that are free from clogging. With a fixed edge grain, the decrease with angle is more rapid. However, notably, at low angles ($\theta<40^\circ$) the flow rate is enhanced significantly in comparison to flat edges. A recent work has also noted an increase in flow rate when using fixed grains at the orifice edges in a horizontal conveyor belt driven flow~\cite{Zhu2019}. These authors attribute the increase to a reduction in the horizontal flow of the grains at the edges that typically impedes the vertical flow of the particles above the orifice.

\begin{figure}
    \centering
    \includegraphics[width=1\linewidth]{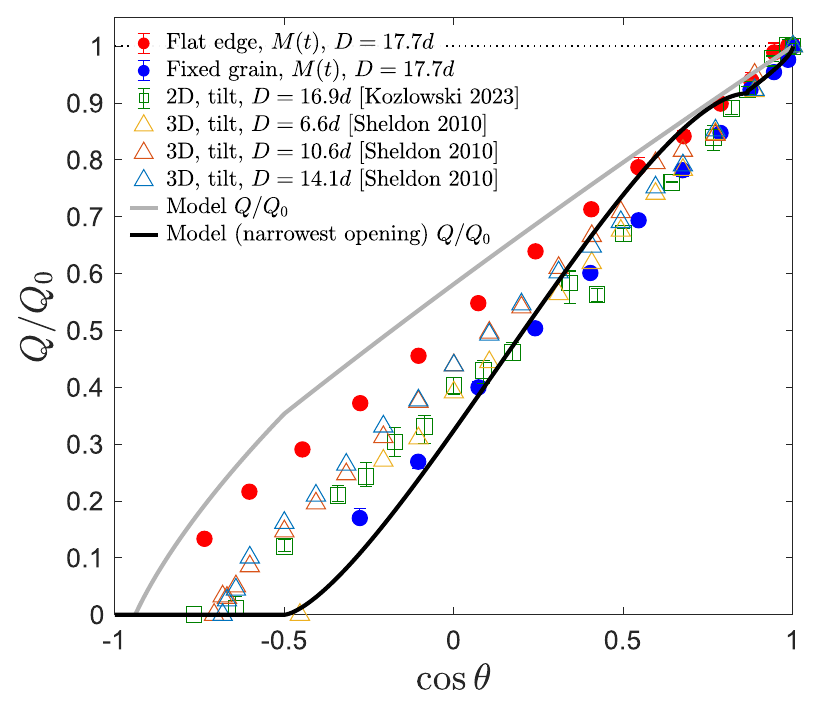}
    \caption{Normalized flow rates $Q/Q_0$ as a function of $\cos(\theta)$. Gray line: our model of flow through the orifice cross-section with $\theta_{\rm S}=60^{\circ}, \theta_{\rm h}=120^{\circ}$ (Sec.~\ref{sec:model}). Black line: our model of flow through narrowest opening cross-section with $\theta_{\rm S}=60^{\circ}, \theta_{\rm h}=120^{\circ}$ (Sec.~\ref{sec:modelnarrowest}). } 
    \label{fig:QQ0}
\end{figure}

The normalized flow rates, defined as $Q/Q(\theta=0^{\circ})=Q/Q_0$, are shown in Fig.~\ref{fig:QQ0} as a function of $\cos\theta$. Prior data from tilted silo experiments of Refs.~\cite{Sheldon2010} and~\cite{Kozlowski2023} are included for reference. The normalized flow rates agree among the data sets for small angles but more clearly differentiate at steeper angles. The flat edge case has consistently faster flows than in the tilted silo experiments, whereas the fixed grain condition has consistently slower flows; each case forms upper and lower bounds, respectively, on flows  when the entire silo is tilted. We will comment on this finding in Sec.~\ref{sec:conclusions}.

\begin{figure}[h!]
    \centering
    \includegraphics[width=1.0\linewidth]{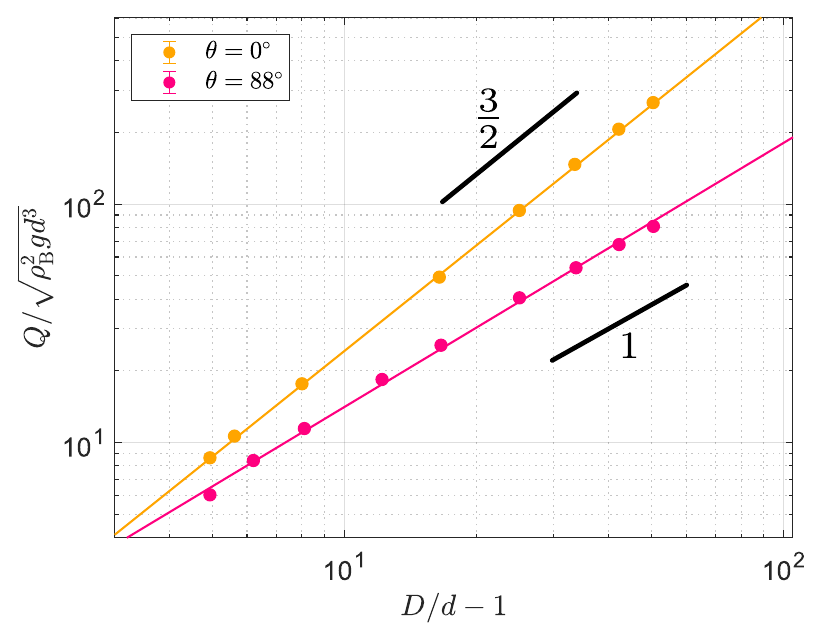}
    \caption{Rescaled flow rate as a function of rescaled orifice size for $\theta=0^{\circ}$ (orange) and $\theta=88^{\circ}$ (pink) in the flat edge case.}
    \label{fig:beverloo}
\end{figure}

\subsection{Beverloo's rule with $\theta\approx 90^{\circ}$}

To compare our system with prior studies of flat-bottomed silos and lateral wall orifices, we next examine the flow rate as a function of orifice diameter $D$ when the orifice is horizontal $\theta=0^{\circ}$ ---the typical configuration for which Eqn.~\ref{eqn:beverloo} applies--- and when the orifice is close to vertical, $\theta=88^{\circ}$. We present results in Fig.~\ref{fig:beverloo}. The scaling for the flat-bottom, conventional silo agrees with the Beverloo's Rule in Eqn.~\ref{eqn:beverloo}, as the power law fit exponent (which Beverloo predicts to be 3/2) on the rescaled axes of Fig.~\ref{fig:beverloo} is $1.48\pm0.02$ (error is $95\%$ confidence interval). However, the exponent is roughly linear when a lateral opening is employed: the fit exponent becomes $1.11\pm 0.03$ (error is $95\%$ confidence interval). 

Though we use a different geometric arrangement in order to tilt our orifice, our findings are consistent with prior research by Zhou et al. on lateral wall orifices in thin, quasi-2D silos (where silo thickness $W\ll D$)~\cite{Zhou2017}. These authors showed, using dimensional analysis and detailed experiments, that when the orifice of a quasi-2D silo is lateral instead of horizontal, $Q$ is \textit{linear} with $D$ but also scales as $W^{1/2}$. The underlying physical mechanism remains unclear, but it is likely that wall friction with the front and back plates plays an important role in modifying the flow pattern of grains as they approach the orifice, slowing the redirection of flow towards the horizontal~\cite{Zou2020}. Further analysis in this direction is outside of the scope of this work.

\begin{figure*}[t]
    \centering
    \includegraphics[width=1.0\linewidth]{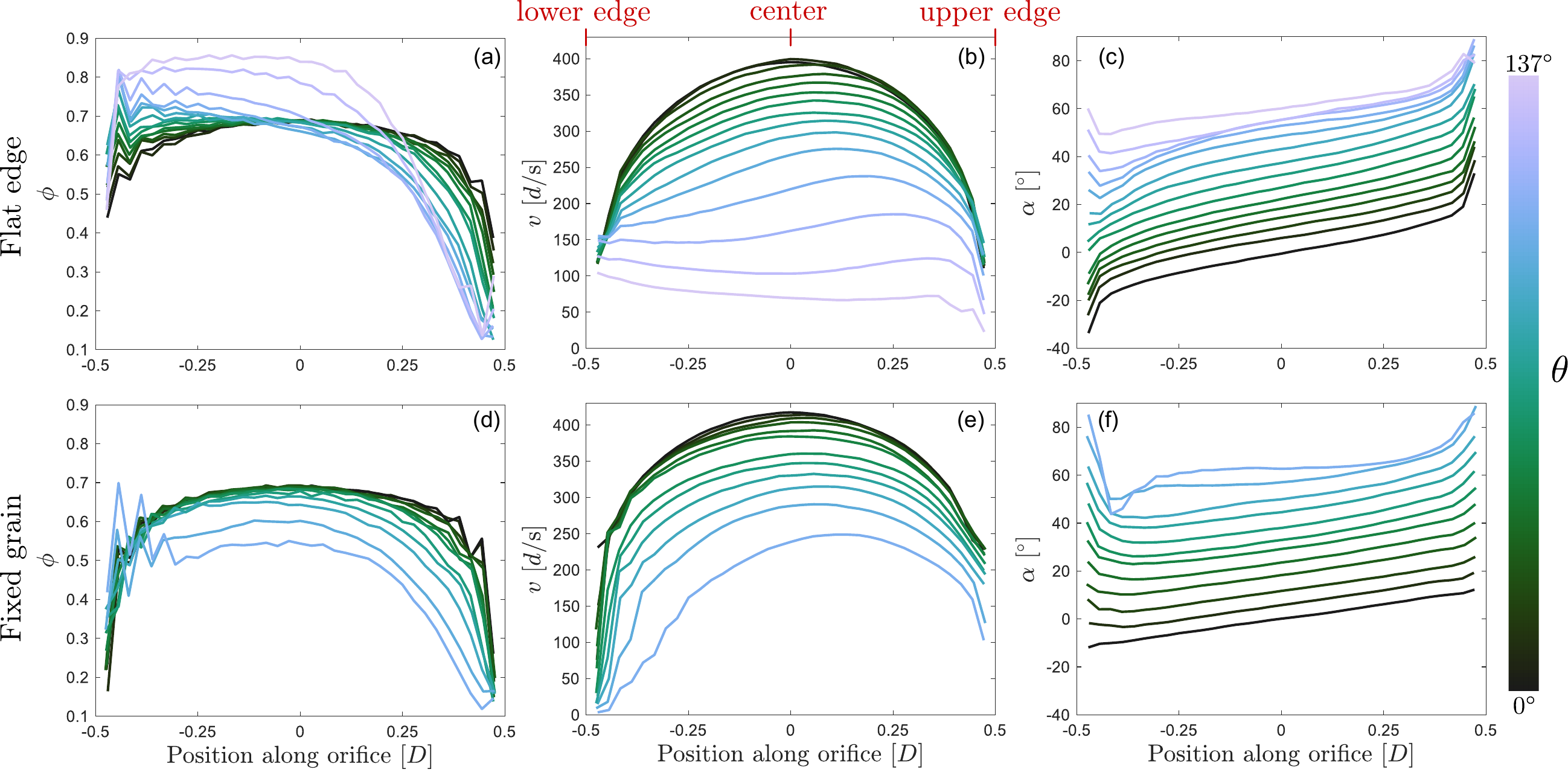}
    \caption{Profiles along the orifice of (a,d) packing fraction, (b,e) speed, and (c,f) flow angle for (top row) the flat edge case and (bottom row) the fixed grain case, at varying elevation angles $\theta$ (see color scale).}
    \label{fig:profiles}
\end{figure*}

\subsection{Velocity and packing fraction at the orifice}\label{sec:profiles}

Using particle tracking as described in Sec.~\ref{sec:expmethods}, we measure the average flow velocity and packing fraction of grains along the orifice profile for varying $\theta$. Figure~\ref{fig:profiles}a-c displays the orifice profiles of packing fraction $\phi$, outpouring speed $v$, and flow angle $\alpha$ for the flat edge case. (See Supplemental Materials for profiles of components $v_{\perp}$ and $v_{||}$~\cite{SupplementalMaterials}.) With increasing angle $\theta$, the material orders in a crystalline lattice more effectively at the lower edge of the orifice, and $\phi$ consequently increases. At the upper edge, in contrast, $\phi$ decreases as the material locally dilates as grains circulate around the upper edge. The speed decreases in magnitude from the center of the orifice towards the edges for $\theta \lesssim 110^\circ$, and an asymmetry develops with the grains that are on the upper half of the orifice flowing faster. Nevertheless, the speed profile is notably more symmetric at all $\theta$ than those observed in tilted silos (see Ref.~\cite{Kozlowski2023}). For $\theta > 110^\circ$, the shape of the speed profile flattens. Importantly, the flow direction $\alpha$ is, except very close to the edges, linear with position along the orifice with a consistent slope across all elevation angles. The only significant change of $\alpha$ with increasing $\theta$ is a constant offset.

With a fixed grain on each orifice edge, the profiles correspondingly change as shown in Fig.~\ref{fig:profiles}d-f, especially near the upper and lower edges. The magnitude of the velocity decreases in a manner similar to that of the flat edge case, but the flow at the lower edge is never as fast as the flow at the upper edge and the profile does not flatten for any $\theta$. The packing fraction shows the most stark differences in our two experimental cases: with the fixed grain, packing fraction across the orifice decreases on average with increasing elevation angle, even at the lower edge of the orifice. We will argue shortly that, in the flat edge case, ordered grains in the nominal stagnant zone creep horizontally along the lower boundary. Thus, the hexagonal, highly packed structure overflows into the orifice. In contrast, with a fixed grain, this order is frustrated as grains flow around the obstructing grain, collectively dilating in the process near the bottom edge. The profile of the direction of flow $\alpha$ is also linear as in the flat edge case, perhaps with smaller deviations close to the edges due to the presence of the fixed grains.

\begin{figure*}[t]
    \centering
    \includegraphics[width=1.0\linewidth]{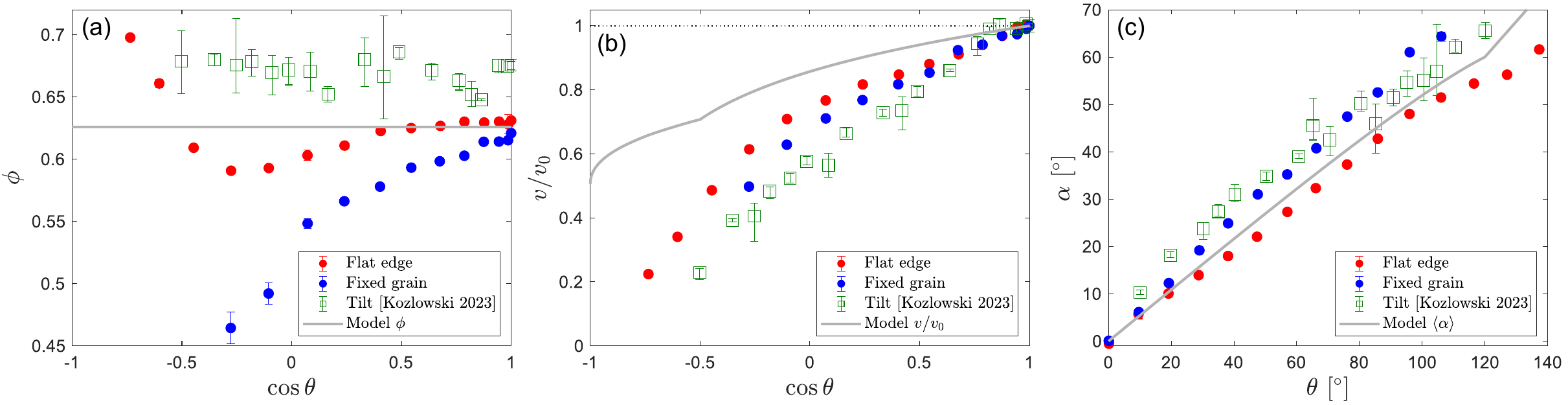}
    \caption{Averages along the orifice profile of (a) packing fraction, (b) normalized outpouring speed, and (c) angle for the flat edge and fixed grain cases. Solid gray lines: Models for each average quantity, discussed in Sec.~\ref{sec:model}. Green data are from the 2D tilted silo experiments of Ref.~\cite{Kozlowski2023}.}
    \label{fig:profileaverages}
\end{figure*}

Figure~\ref{fig:profileaverages} displays averages taken along the orifice profiles of Fig.~\ref{fig:profiles} of packing fraction, outpouring speed $v/v_0$ normalized to the average speed at $\theta = 0^{\circ}$, and angle $\alpha$ as a function of $\theta$. Packing fraction changes nonmonotonically as elevation angle increases with a flat edge, but decreases monotonically with a fixed grain. The speed linearly decreases as $\cos\theta$ decreases from 1 in both cases until around $\theta\approx 90^{\circ}$, beyond which the speed declines more rapidly. The decline is stronger for the fixed grain case. As noted above, the outpour angle $\alpha$ linearly increases with elevation angle in both cases, albeit with a different slope, and in the flat edge case there is an inflection point around $120^{\circ}$. 

On each of the plots shown, we also include corresponding results from Ref.~\cite{Kozlowski2023}, in which particle-scale flow was analyzed for a slightly smaller orifice size of $16.9d$ with a tilted silo. (This is the only tilted-silo study, to our knowledge, which captured particle-scale motion.) Unlike in the current configuration, the packing fraction for the tilted silo is constant with angle. The normalized speed, however, is comparable between the tilted silo and fixed grain case (and is significantly lower for a given angle than in the flat edges case). Moreover, the outpouring angle for the tilted silo, though greater in magnitude at small angles, is approximately the same as for the fixed grain case.

\subsection{Flow in the vicinity of the orifice}\label{sec:spatialflow}

Having characterized the flow across the orifice line, we lastly turn to the flow of the material in the vicinity of the orifice. This information will be vital to modeling the average exit flow characteristics of Fig.~\ref{fig:profileaverages}, which we will describe in Sec.~\ref{sec:model}. 

\begin{figure*}
    \centering
    \includegraphics[width=1\linewidth]{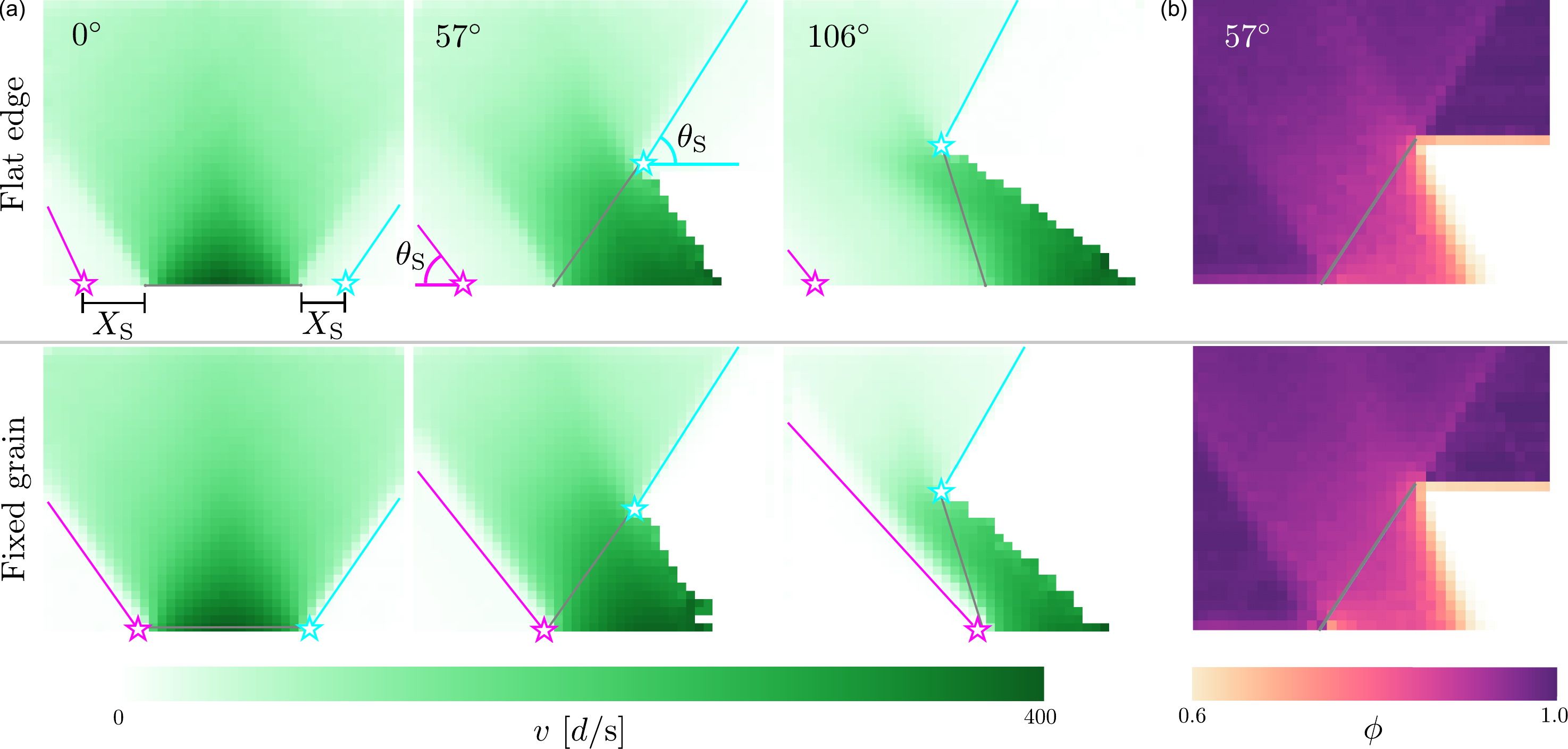}
    \caption{(a) Heat maps for three sample angles $\theta$ showing average grain speed $v$ sampled across the entire image region for (top row) flat edge and (bottom row) fixed grain. Gray line: orifice. Stars: edge of lower (pink) and upper (cyan) stagnant zone, each a horizontal distance $X_{\rm S}$ from the orifice edge. Lines: Best-fit line to lower (pink) and upper (cyan) stagnant zone boundary, each giving stagnant zone angle $\theta_{\rm S}$. (b) Heat maps of packing fraction at $\theta=57^{\circ}$ for each experimental case.}
    \label{fig:heatmaps}
\end{figure*}

To quantify average spatial flow velocity, we sample the average velocity of grains passing through boxes of size $d\times d$ across the entire image. The averaging protocol is identical to that described in Sec.~\ref{sec:expmethods} for the orifice profile, though in this case we sample every 100 frames in a given trial for temporal averaging. Representative speed heat maps obtained in this manner are shown in  Fig.~\ref{fig:heatmaps}a for a few $\theta$ in both orifice edge geometries. There are clear stagnant zones on or near each orifice edge, evidenced by very low speeds. Moreover, as shown with representative packing fraction heat maps in Fig.~\ref{fig:heatmaps}b, the stagnant regions have densely packed grains whereas the material dilates in the fast flow regions; recall that we use monodisperse grains, which can pack in 2D most efficiently to $\phi_{\rm max}\approx 0.91$. (See the Supplemental Materials for heat maps of all of the experimental conditions~\cite{SupplementalMaterials}.)

\subsubsection{Stagnant zones}\label{sec:staggeometry}

\begin{figure}
    \centering
    \includegraphics[width=1\linewidth]{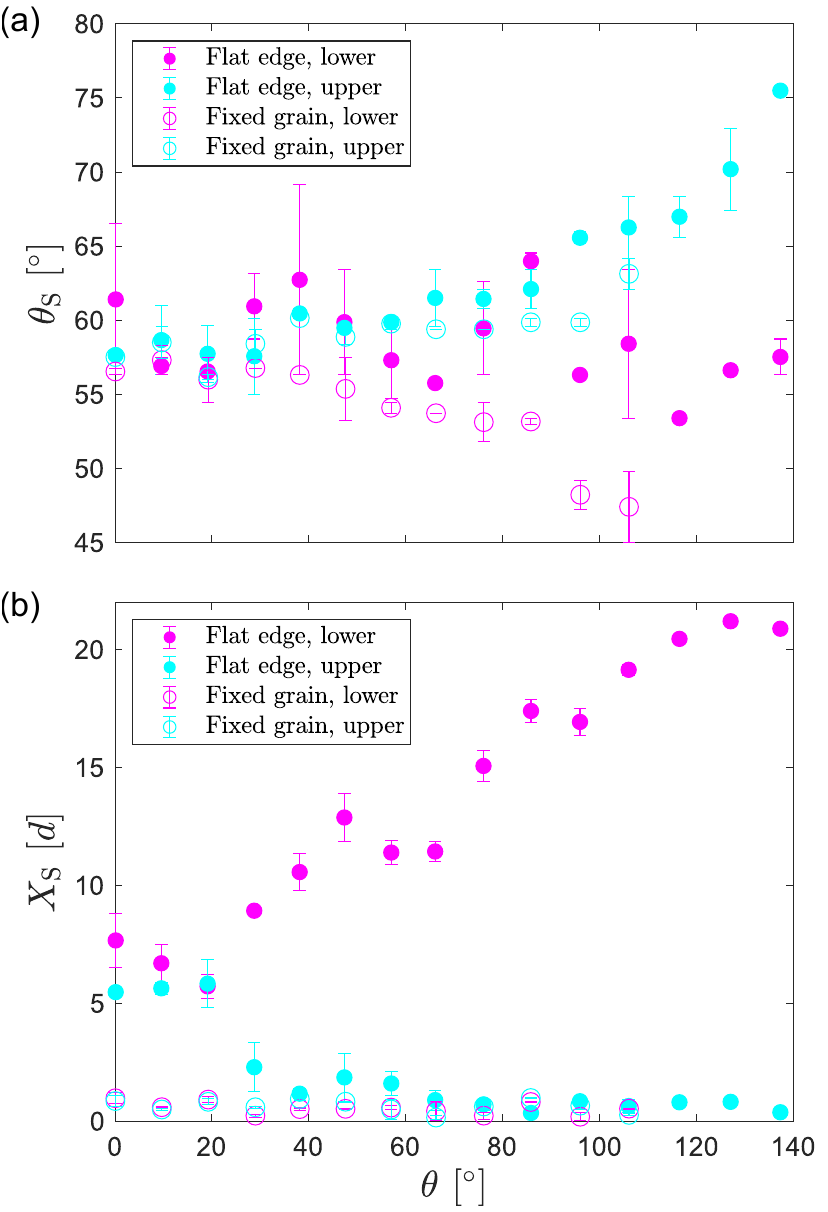}
    \caption{(a) Stagnant zone angles $\theta_{\rm S}$ and (b) edge distances $X_{\rm S}$ measured at the lower (pink) and upper (cyan) edges for the flat edge case (filled circles) and fixed grain case (open circles).}
    \label{fig:stagangles}
\end{figure}

Two important features of each stagnant zone are the stagnant zone angle $\theta_{\rm S}$ from the horizontal and the horizontal distance $X_{\rm S}$ from the stagnant zone to the orifice edge. We define the stagnant zone regions at the lower and upper orifices by thresholding the speed heat map to $5\%$ of the maximum speed per heat map. (The exact choice of a relative threshold or absolute speed threshold does not influence our results.) $\theta_{\rm S}$ is then the angle of the best-fit line to the sloped edge of a detected stagnant zone region; see pink and cyan lines in Fig.~\ref{fig:heatmaps}a. $X_{\rm S}$ is the minimum horizontal distance from the stagnant zone region to the orifice edge. We determine $\theta_{\rm S}$ and $X_{\rm S}$ for the lower orifice and the upper orifice separately, as shown in Fig.~\ref{fig:stagangles}a,b. 

For both the flat edge and fixed grain cases, the stagnant zone angle is close to the expected $60^{\circ}$ for hexagonal ordering across angles $\theta$ with some deviation at the steepest angles. We will therefore set $\theta_{\rm S}=60^{\circ}$ in the models we establish in Secs.~\ref{sec:model} and~\ref{sec:modelnarrowest}. A more prominent difference between the flat edge and fixed grain cases, though, is how far from the orifice edges the stagnant zone edges are. For the fixed grain case, the stagnant zone is directly at the orifice edge due to the constraint imposed by the glued bead. However, for the flat edge case, the development of $X_{\rm S}$ with $\theta$ is not trivial. At the lower edge, $X_{\rm S}$ increases approximately linearly with $\theta$, indicating an increasing size of a slip zone~\cite{Zou2020}. At the upper edge, beyond $\theta\approx 30^{\circ}$, the slip zone vanishes. These observations, along with visual confirmation from Fig.~\ref{fig:heatmaps}, indicate that as $\theta$ increases, a channel of ``horizontal flow'' forms along the base next to the lower orifice boundary, since grains can slide collectively along this base. Adding a fixed grain prevents this channel from forming altogether, and therefore the maximum achievable $\theta$ is significantly smaller in that case.

\subsubsection{Stagnant zone creep}\label{sec:stagcreep}

\begin{figure*}
    \centering
    \includegraphics[width=1\linewidth]{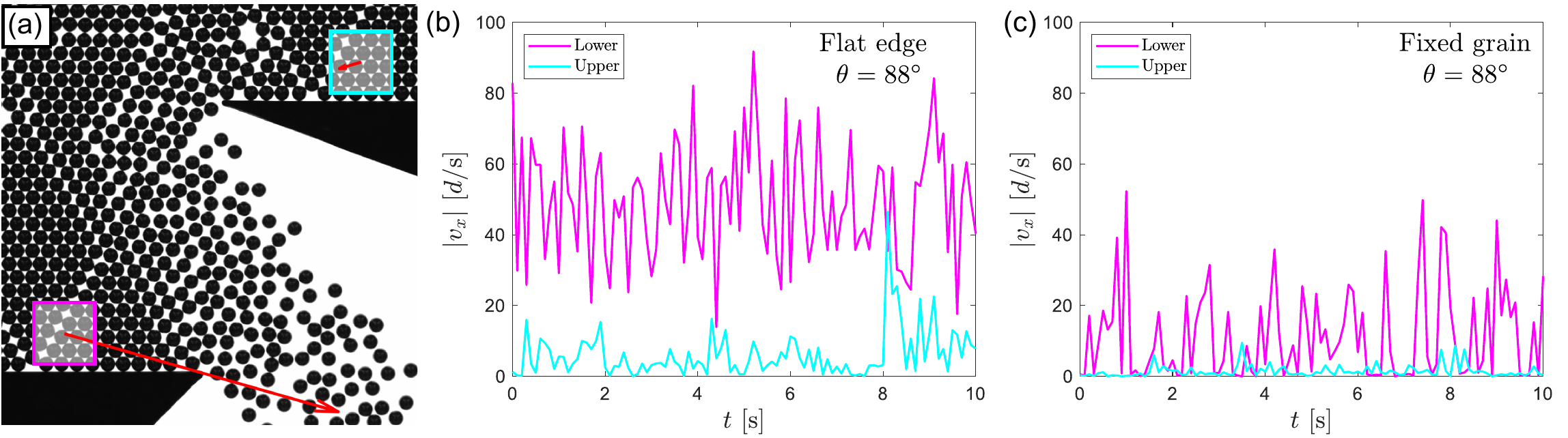}
    \caption{(a) Sample image showing the box region used to obtain the average velocity of the ``stagnant'' zone; the sampling boxes are $4d\times 4d$ and centered $9d$ from the orifice edges. Horizontal creep speed as a function of time for (a) the flat edge case and (b) the fixed grain case at $\theta = 88^{\circ}$. }
    \label{fig:stagcreep1}
\end{figure*}

Motivated by our observations of channel formation in the flat edge case, we next quantify the average horizontal speed of grains in the (nominal) stagnant zone. Figure~\ref{fig:stagcreep1}a shows the two sampling regions used to determine horizontal creep: $4d\times 4d$ boxes at the lower and upper orifices, each centered $9d$ away from their respective orifice edges. The qualitative results we show do not depend on exact sampling size or box position.

Horizontal speeds $|v_x(t)|$ time series for each region with $\theta=88^{\circ}$ are shown in Fig.~\ref{fig:stagcreep1}b,c. In some cases, especially small angles and with a fixed grain, the material in the stagnant zone flows intermittently. With flat edges at steep angles, however, flow at the lower edge is continuous with fluctuations as grains near the orifice edge compete with non-stagnant zone grains to exit the orifice. Though we do not explore this further in the current work, the competition of horizontally flowing grains is a plausible explanation for the increase in flow rate at small angles with the addition a fixed grain as proposed in \cite{Zhu2019}.

\begin{figure}
    \centering
    \includegraphics[width=1\linewidth]{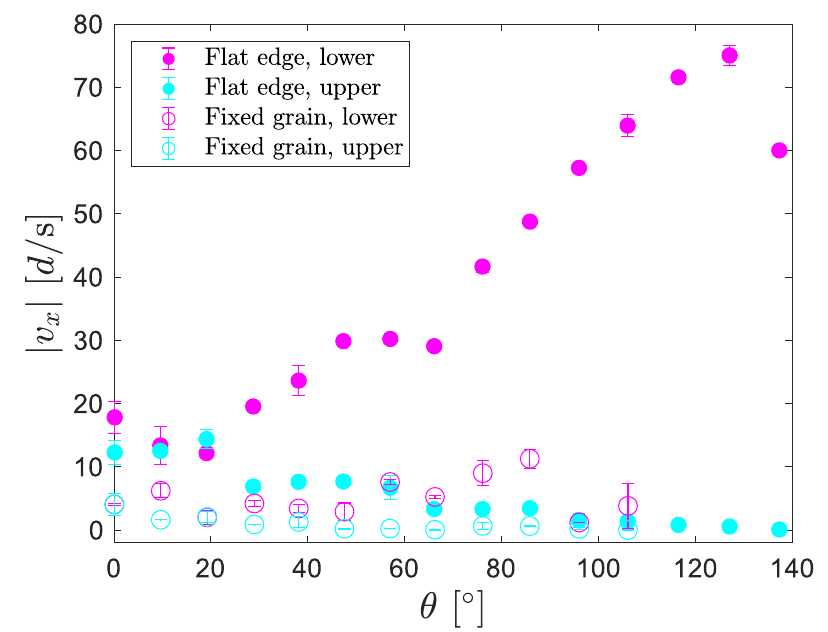}
    \caption{Magnitude of average horizontal stagnant zone creep velocity at the lower (pink) and upper (cyan) edges for the flat edge case (filled circles) and fixed grain case (open circles).}
    \label{fig:stagcreep2}
\end{figure}

We lastly highlight in Fig.~\ref{fig:stagcreep2} the time average $|v_x|$ for both experimental cases. In the flat edge case, at the lower edge the material flows more rapidly in the horizontal direction as $\theta$ increases and the slip zone widens. In fact, horizontal creep becomes the most prominent contribution of the flow near the lower orifice edge. The average flow speed near the lower orifice edge in the profiles shown in Fig.~\ref{fig:profiles}b,e is $\approx 100 d/$s. 

\section{Model for $Q(\theta)$}\label{sec:model}

Based on our observations of the nontrivial dependence of mass discharge rate $Q$ with orifice angle $\theta$, and using our findings for the average flow properties at the orifice from particle tracking analysis,  we develop a physical model of $Q(\theta)$. The flow rate through an orifice of width $D$ at angle $\theta$ with the horizontal is 
\begin{equation}\label{eqn:qmodel}
    Q = D \rho_0(\theta) v(\theta) \cos[\alpha(\theta)]
\end{equation}
where the material has an average exit speed $v$, average angle $\alpha$ with respect to the orifice normal, and average density $\rho_0$ at the orifice. A model for $Q/Q_0$ therefore requires modeling of $\rho_0(\theta)$, $v(\theta)$, and $\alpha(\theta)$.

\subsection{$\alpha(\theta)$}

\begin{figure}
    \centering
    \includegraphics[width=1.0\linewidth]{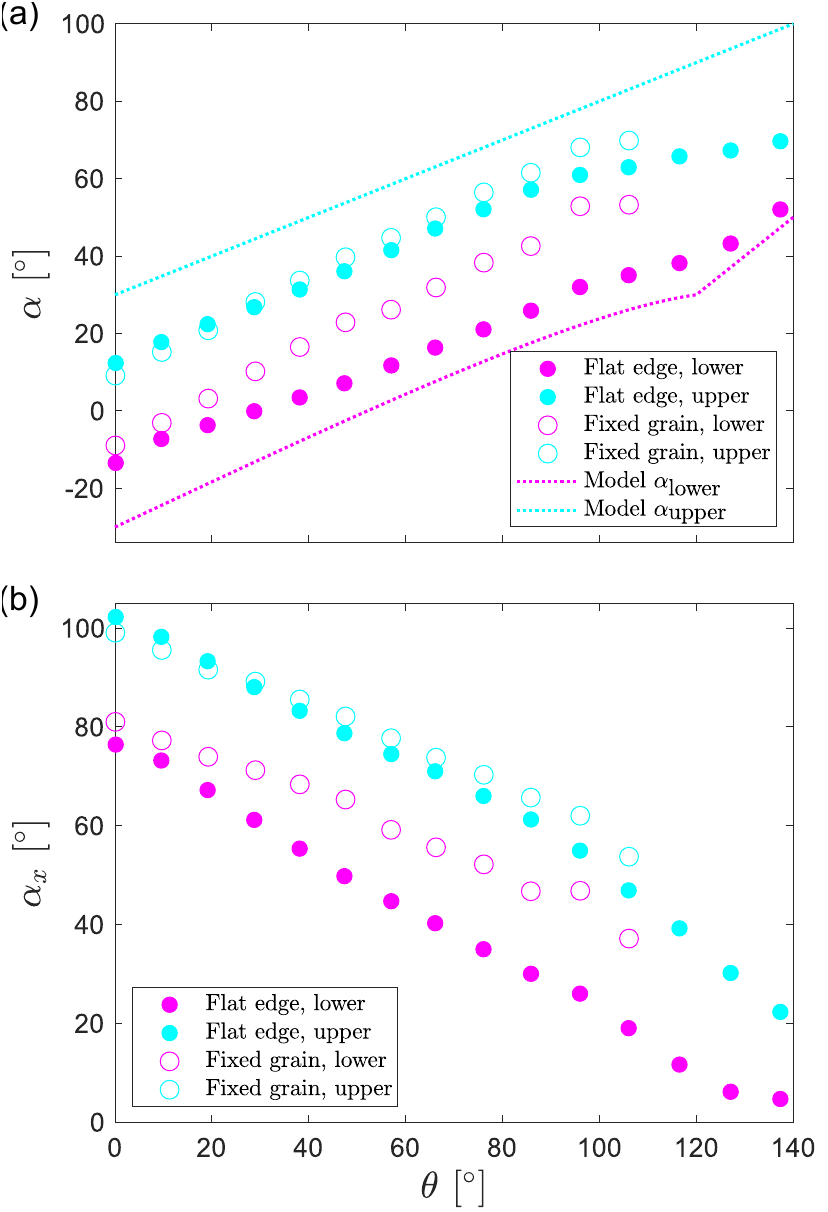}
    \caption{(a) $\alpha_{\rm lower}$ and $\alpha_{\rm upper}$ from the profiles in Fig.~\ref{fig:profiles}c sampled $2d$ inward from the orifice edges. Dashed lines: models for lower and upper angles. (b) Flow with respect to the horizontal $\alpha_{x,\rm lower}$ and $\alpha_{x,\rm upper}$ from the profiles in Fig.~\ref{fig:profiles}c sampled $2d$ inward from the orifice edges.}
    \label{fig:averagealpha}
\end{figure}

We showed in Sec.~\ref{sec:profiles} that $\alpha$ is linear across the orifice. This implies that the average flow direction can be calculated from the flow direction at each edge of the orifice $\alpha_{\rm lower}$ and $\alpha_{\rm upper}$:
\begin{equation}\label{eqn:averagealphamodel}
    \alpha = \left<\alpha\right> = (\alpha_{\rm lower} + \alpha_{\rm upper})/2.
\end{equation}
Modeling the direction of flow close to the orifice edges should be simpler since the boundaries constraint the flow. In Fig.~\ref{fig:averagealpha}a we display  the flow angles $\alpha_{\rm lower}$ and $\alpha_{\rm upper}$ from the profiles shown in Fig.~\ref{fig:profiles}c a distance $2d$ inward from the lower and upper orifice edges to avoid boundary measurement issues in the bins directly next to each edge. Recall that we define $\alpha>0$ for exit velocities that are below (clockwise from) the orifice normal.

\subsubsection{$\alpha_{\rm lower}(\theta)$}

\begin{figure*}
    \centering
    \includegraphics[width=1.0\linewidth]{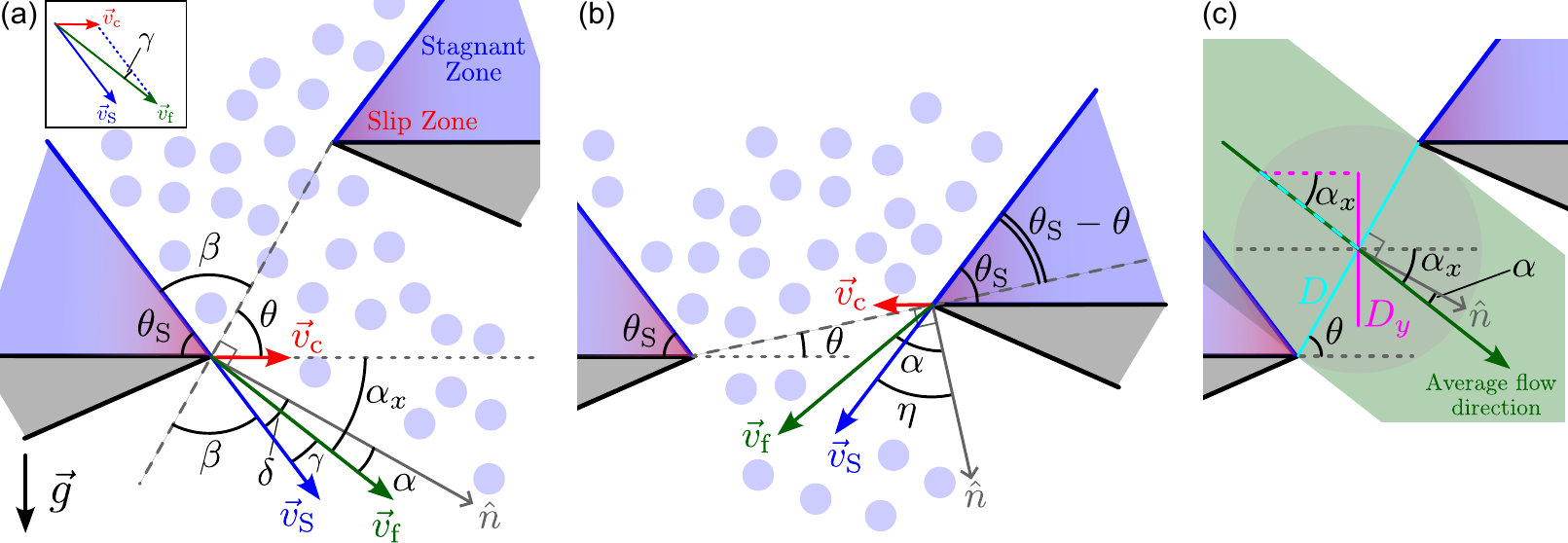}
    \caption{(a) Calculation of angle $\alpha$ between flow direction $\vec{v}_{\rm f}$ and normal $\hat{n}$ to the orifice at angle $\theta$ at the lower orifice edge. $\vec{v}_{\rm c}$ is the average creep velocity at the bottom edge, $\vec{v}_{\rm S}$ is the average flow of the material along the bottom edge stagnant zone ($\theta_{\rm S}\approx 60^{\circ}$), and $\vec{v}_{\rm f}$ is the net velocity at the bottom edge. $\beta$ is the angle between the orifice line and $\vec{v}_{\rm S}$, and $\delta$ is the angle $\vec{v}_{\rm S}$ makes with respect to $\hat{n}$. (b) Calculation of angle $\alpha$ between flow direction $\vec{v}_{\rm f}$ and normal $\hat{n}$ to the orifice at angle $\theta$ at the upper orifice edge. $\eta$ is the angle $\vec{v}_{\rm S}$ makes with respect to $\hat{n}$. (c) Schematic of orifice for determining the free-fall lengthscale of the material based on average flow of material at the orifice. $D$ is the orifice size,  $D_y$ is the vertical free-fall lengthscale, and $\alpha_x$ is the angle of the flow direction with respect to the horizontal.  }
    \label{fig:modelschematics}
\end{figure*}

To model $\alpha_{\rm lower}$, we first assume that the net flow velocity $\vec{v}_{\rm f} \equiv \vec{v}_{\rm S}+ \vec{v}_{\rm c}$ at the lower edge is determined by a contribution of velocity along the stagnant zone boundary $\vec{v}_{\rm S}$ and average horizontal creep $\vec{v}_{\rm c}$, as shown in Fig.~\ref{fig:modelschematics}a. At small orifice angles $\theta$, horizontal creep is a relatively small contribution to the flow, so $\vec{v}_{\rm f}\approx \vec{v}_{\rm S}$. However, we have shown (Fig.~\ref{fig:stagcreep2}) that the average creep speed $v_{\rm c}$ increases proportionally with $\theta$, suggesting that at large enough $\theta$, the flow at the lower edge (in the flat edge case) becomes more horizontal. We demonstrate this to be true in Fig.~\ref{fig:averagealpha}b, which shows the lower and upper edge flow angles with respect to the horizontal, $\alpha_x = \pi/2-\theta+\alpha$ (see Fig.~\ref{fig:modelschematics}a). For the flat edge case, at the lower orifice edge the angle approaches $0^{\circ}$ around $\theta=120^{\circ}$, indicating primarily horizontal flow contributed by the stagnant zone creep.  Given these observations regarding stagnant zone creep, we thus model the creep speed as

\begin{equation*}
   v_c =  \left\lbrace \begin{array}{ll}
\theta v_{\rm f} /\theta_{\rm h} , & \theta \leq \theta_{\rm h}\\
v_{\rm f} , & \theta>\theta_{\rm h}
\end{array} \right.,
\end{equation*}
where $\theta_{\rm h}$ is a threshold angle beyond which the flow at the lower edge is purely horizontal. We will set $\theta_{\rm h}=120^{\circ}$ based on the results shown in Fig.~\ref{fig:averagealpha}b.

We next focus on the geometry shown in Fig.~\ref{fig:modelschematics}a to determine the angle $\alpha$ between $\vec{v}_{\rm f}$ and $\hat{n}$. Define the angle $\beta = \pi - \theta_{\rm S} - \theta$ between $\vec{v}_{\rm S}$ and the orifice line. Then the angle between $\vec{v}_{\rm S}$ and the orifice normal is $\delta=\pi/2-\beta = \theta + \theta_{\rm S} - \pi/2$. As shown in the upper inset of Fig.~\ref{fig:modelschematics}a, 

\begin{equation*}
    \frac{\sin(\pi-\theta_{\rm S})}{v_{\rm f}} = \frac{\sin\gamma}{v_{\rm c}} = \left\lbrace \begin{array}{ll}
\frac{\theta_{\rm h}}{\theta v_{\rm f} } \sin\gamma , & \theta\leq\theta_{\rm h}\\
\frac{1}{v_{\rm f}} \sin\gamma , & \theta >\theta_{\rm h}
\end{array} \right. .
\end{equation*}
Hence
\begin{equation*}
    \gamma =  \left\lbrace \begin{array}{ll}
\sin^{-1} \left( \frac{\theta}{\theta_{\rm h}}\sin\theta_{\rm S}\right) , & \theta\leq\theta_{\rm h}\\
\theta_{\rm S} , & \theta >\theta_{\rm h}
\end{array} \right. ,
\end{equation*}
and given that $\alpha_{\rm lower}(\theta) = \delta-\gamma$, 

\begin{multline}\label{eqn:modelalphalower}
        \alpha_{\rm lower}(\theta) \\= \left\lbrace \begin{array}{ll}
            \theta + \theta_{\rm S} - \frac{\pi}{2} - \sin^{-1} \left( \frac{\theta}{\theta_{\rm h}}\sin\theta_{\rm S}\right) , & \theta\leq\theta_{\rm h}\\
            \theta - \frac{\pi}{2} , & \theta >\theta_{\rm h}
        \end{array} \right.
\end{multline}
    

Equation~\ref{eqn:modelalphalower} is shown as a pink dashed line in Fig.~\ref{fig:averagealpha}a. We set $\theta_{\rm S}=60^\circ$ based on the measurements of stagnant zone angle discussed in Sec.~\ref{sec:staggeometry} and $\theta_{\rm h}=120^\circ$ based on the increasing importance of horizontal creep for the flat edge case shown in Figs.~\ref{fig:stagcreep2} and~\ref{fig:averagealpha}b. The model captures the slope for the flat edge case even though it is shifted by an additive constant with respect to the experimental data. Recall that $\alpha_{\rm lower}$ was measured in practice two particles away from the edge of the orifice.) This model does not apply to the fixed grain case given that horizontal creep is strongly reduced in this case.

\subsubsection{$\alpha_{\rm{upper}}(\theta)$}

We now turn to $\alpha_{\rm upper}$. When $\theta$ is close to $0^{\circ}$, the direction of flow next to the upper edge of the orifice is essentially the direction imposed by the upper stagnant zone. In this case, $\alpha\approx\eta(\theta)=\theta-\theta_{\rm S}+\pi/2$, where the angle $\eta$ is defined in Fig.~\ref{fig:modelschematics}b.  When $\theta$ is large, however, the direction of flow next to the upper edge of the orifice is essentially the direction imposed by the lower stagnant zone. Therefore, $\alpha\approx\delta(\theta)=\theta+\theta_{\rm S}-\pi/2$,  ($\delta$ is defined in Fig.~\ref{fig:modelschematics}a).

Having a clear physical picture for small and large $\theta$ limits, we propose a simple linear interpolation:

\begin{equation}\label{eqn:modelalphaupper}
    \alpha_{\rm upper}(\theta) = \eta(\theta) (1-\theta/\theta_{\rm h})+\delta(\theta) \theta/\theta_{\rm h}.
\end{equation}

The prediction of this model with $\theta_{\rm S}=60^\circ$ and  $\theta_{\rm h}=120^\circ$ is shown as a cyan dashed line in Fig.~\ref{fig:averagealpha}a. The agreement with the data in terms of slope is fair with the exception of the very large angles $\theta > \theta_{\rm h}$ where the contribution of the creep may be important. Again, the model offsets the data by an additive constant as we observed with $\alpha_{\rm lower}$, mainly because the measurements were taken two particles away from the edge.

Lastly, we combine the models for $\alpha_{\rm lower}$ and $\alpha_{\rm upper}$ with Eqn.~\ref{eqn:averagealphamodel} for $\alpha$ and ultimately obtain

\begin{multline}\label{eqn:modelalphacomplete}
    \alpha(\theta) \\=  \left\lbrace 
    \begin{array}{ll}    
        \theta\big(1+\frac{\theta_{\rm S}-\pi/2}{\theta_{\rm h}}\big)-\frac{1}{2}\sin^{-1}\big(\frac{\theta}{\theta_{\rm h}}\sin(\theta_{\rm S})\big), & \theta \leq \theta_{\rm h}\\
        \theta\big(1+\frac{\theta_{\rm S}-\pi/2}{\theta_{\rm h}}\big)-\frac{\theta_{\rm S}}{2}, & \theta > \theta_{\rm h}
    \end{array} \right.
\end{multline}

Figure~\ref{fig:profileaverages}c displays this model as a gray line using $\theta_{\rm S}=60^{\circ}$ and $\theta_{\rm h}=120^{\circ}$, both of which are physically motivated values. The model agrees well with the flat edge data both quantitatively and qualitatively, capturing the curvature of the data just before $120^{\circ}$ and the upward trend beyond this angle.

\subsection{$v(\theta)$}

We next turn to the average discharge speed of the material. The material accelerates in a limited region in the vicinity of the orifice \cite{Janda2012,RubioLargo2015,Darias2020}, which we model as a zone where the material dilates enough for grains to fall under gravity without dissipation (a so-called \textit{free-fall arch} with characteristic lengthscale $R$). For upright silos with a horizontal orifice, this simplified model sets $R\propto D$ based on a roughly semicircular free-fall region above the orifice, and therefore $v\propto{\sqrt{gD}}$. For a vertical orifice on a side wall where the average flow direction is not purely vertical, Zou, et. al., showed in Ref.~\cite{Zou2020} that  $R\propto D\sin{\alpha_x}$ describes the flow rate of the material. $\alpha_x$ is the average direction of flow at the orifice with respect to the horizontal. We adopt Zou's approach as a first-order model for speed in our system as well. As shown in Fig.~\ref{fig:modelschematics}a,c, $\alpha_x = \pi/2-\theta+\alpha$; hence, we define the vertical free-fall lengthscale $R = D_y = D\sin{\alpha_x}$ and $v\propto \sqrt{D_y}$. Upon normalizing by the speed $v_0$ at $\theta=0$, we obtain 

\begin{equation}\label{eqn:modelspeed}
    v/v_0= \sqrt{\cos(\theta-\alpha)}.
\end{equation}

Figure~\ref{fig:profileaverages}b shows the experimentally measured normalized speeds $v/v_0$ as well as the prediction from Eqn.~\ref{eqn:modelspeed}, using $\alpha$ from Eqn.~\ref{eqn:modelalphacomplete}. Although the model for $\alpha(\theta)$ is fair (Fig.~\ref{fig:profileaverages}c), this approach is not adequate to describe the dependence of speed on orifice angle, overestimating the speed for all except small angles $\theta$ where $\alpha\approx\theta$ and $\cos(\theta-\alpha)\approx 1$. Despite the poor agreement in speed, we test in the following how this impacts the prediction of the flow rate.

\subsection{$\rho_0(\theta)$}

The average density of the granular material at the orifice is proportional to average packing fraction $\phi$ by grain mass. Recall from Fig.~\ref{fig:profileaverages}a that the average packing fraction $\phi$ along the orifice is approximately constant for the tilted silo experiment from Ref.~\cite{Kozlowski2023}. It is not constant in the current experiments; however, the changes in $\phi$ are second-order in comparison to those of $v$ and $\cos(\alpha)$, so we model a constant $\phi(\theta)= 0.63$, the average orifice packing fraction for the flat edge case. Therefore $\rho_0(\theta) = \phi \rho = 4.9\;{\rm kg/m}^2$. 

\subsection{Complete model for the flow rate}

Combining the model equations for $\alpha$ (Eqn.~\ref{eqn:modelalphacomplete}), $v/v_0$ (Eqn.~\ref{eqn:modelspeed}), and $\rho_0={\rm constant}$ into Eqn.~\ref{eqn:qmodel}, and setting $\theta_{\rm S}=60^{\circ}$ and $\theta_{\rm h}=120^{\circ}$, we obtain the flow rate model result shown as a gray solid line in Fig.~\ref{fig:QQ0}. We emphasize that the selections of  $\theta_{\rm S}$ and $\theta_{\rm h}$ are physically motivated based on measurements from the experiments. The model prediction captures the trend for the flat edge case in our experiment, but it overestimates the flow rate. We remind readers that the model for $\alpha$ (Eqn.~\ref{eqn:modelalphacomplete}) is most appropriate for modeling the flat edge, not the fixed edge case, since horizontal creep is always insignificant for the fixed grain case (Sec.~\ref{sec:stagcreep}); we therefore do not expect strong agreement with the fixed grain data.

Modeling flow by analyzing flow properties on the putative orifice cross-section, although the most common approach in horizontal orifices~\cite{Janda2012}, may not be the most appropriate method for tilted orifices. In fact, other works have focused on other cross-sections, such as a semicircular arc around the orifice opening in Refs.~\cite{huang2006relationship,Li2022}. Below, we consider an alternative flow cross-section, and find that modeling flow in the modified cross-section is simpler and agrees more robustly with the experimental data. 

\section{Improved model for $Q(\theta)$: narrowest opening cross-section}\label{sec:modelnarrowest}

\begin{figure}
    \centering
    \includegraphics[width=1.0\linewidth]{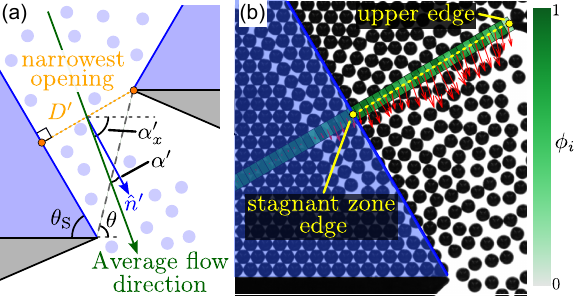}
    \caption{(a) Schematic showing the narrowest opening for $\theta>\pi/2-\theta_{\rm S}$. (b) A sample image in which the narrowest opening cross-section profile has been sampled. Though the cross-section extends to the end of the image, we restrict our analysis to the narrowest opening, between the stagnant zone edge and upper edge.}
    \label{fig:neworificemodel}
\end{figure}

We have shown in the previous section that flow properties along the orifice cross-section are not trivial to model and do not lead to a strong match with experimentally measured flow rates. We now consider an alternative cross-section: the narrowest opening cross-section that the material passes through within the silo. We model the material as having fixed boundaries set by the stagnant zones at constant angle $\theta_{\rm S}=60^{\circ}$ at each orifice edge. In practice this corresponds to neglecting the creep $v_{\rm c}$ of the stagnant zones. For $\theta>\pi/2-\theta_{\rm S} = 30^{\circ}$, as demonstrated in Fig.~\ref{fig:neworificemodel}a, the narrowest opening that grains pass through is not the actual orifice, but (upstream from the orifice) the shortest line connecting the stagnant zone boundary to the upper orifice edge. The length $D'(\theta)$ of this cross-section is

\begin{equation}\label{eqn:Dprime}
    D'(\theta) = \left\lbrace \begin{array}{ll}
    D , & \theta \leq \pi/2- \theta_{\rm S}\\
    D\sin(\theta+\theta_{\rm S}) , & \theta>\pi/2 - \theta_{\rm S}
    \end{array}\right. .
\end{equation}
We note that this approach assumes that the flow drops to zero if $\theta + \theta_{\rm S} = \pi$, i.e., $\theta=120^\circ$ in our case. This assumption is most appropriate for the fixed grain case and less applicable in the flat edge case.

We now analyze the experimental image data by capturing profiles of packing fraction $\phi'$, speed $v'$, and flow angle $\alpha'$ along the narrowest opening cross-section, \textit{not} the orifice line. The flow rate from this new cross-section is determined by

\begin{equation}\label{eqn:qmodelnarrow}
    Q = D'(\theta) \rho_0'(\theta) v'(\theta) \cos[\alpha'(\theta)] ,
\end{equation}

which will lead to the same result as Eqn.~\ref{eqn:qmodel} by mass conservation, except for the fact that the profile we analyze below is truncated to extend only along $D'$, from the stagnant zone boundary to the upper orifice edge. 

Despite the truncated cross-section, we have confirmed that the most significant contributions to the outpouring flow that we measure are indeed captured within the camera's field of view. In Fig.~\ref{fig:Q}, we show the flow rate measured via particle tracking (see Eqn.~\ref{eqn:qvideo}) using the traditional orifice cross-section $D$ (diamonds), using the cross-section along the narrowest opening and extending  across the entire image field of view as shown by the dark green strip in Fig.~\ref{fig:neworificemodel}b (squares), and using the ``truncated'' narrowest opening profile of length $D'$ as shown by the yellow dashed line in Fig.~\ref{fig:neworificemodel}b (stars). Even with truncation of the profile, the flow rates in the fixed grain case are hardly changed. For flat edges there is more significant reduction in the measured flow rate in intermediate angles due to the significant slip zone discussed in Sec.~\ref{sec:spatialflow}. We therefore continue our analysis using the truncated narrowest opening profile specified by the model, extending from the stagnant zone boundary to the upper edge and of length $D'$. 

Before proceeding, we note that the flow cannot be analyzed for $\theta>120^{\circ}$ in this manner since $D^{\prime}$ vanishes at $120^{\circ}$.

\subsection{Flow velocity and packing fraction along the narrowest opening}

\begin{figure*}
    \centering
    \includegraphics[width=1.0\linewidth]{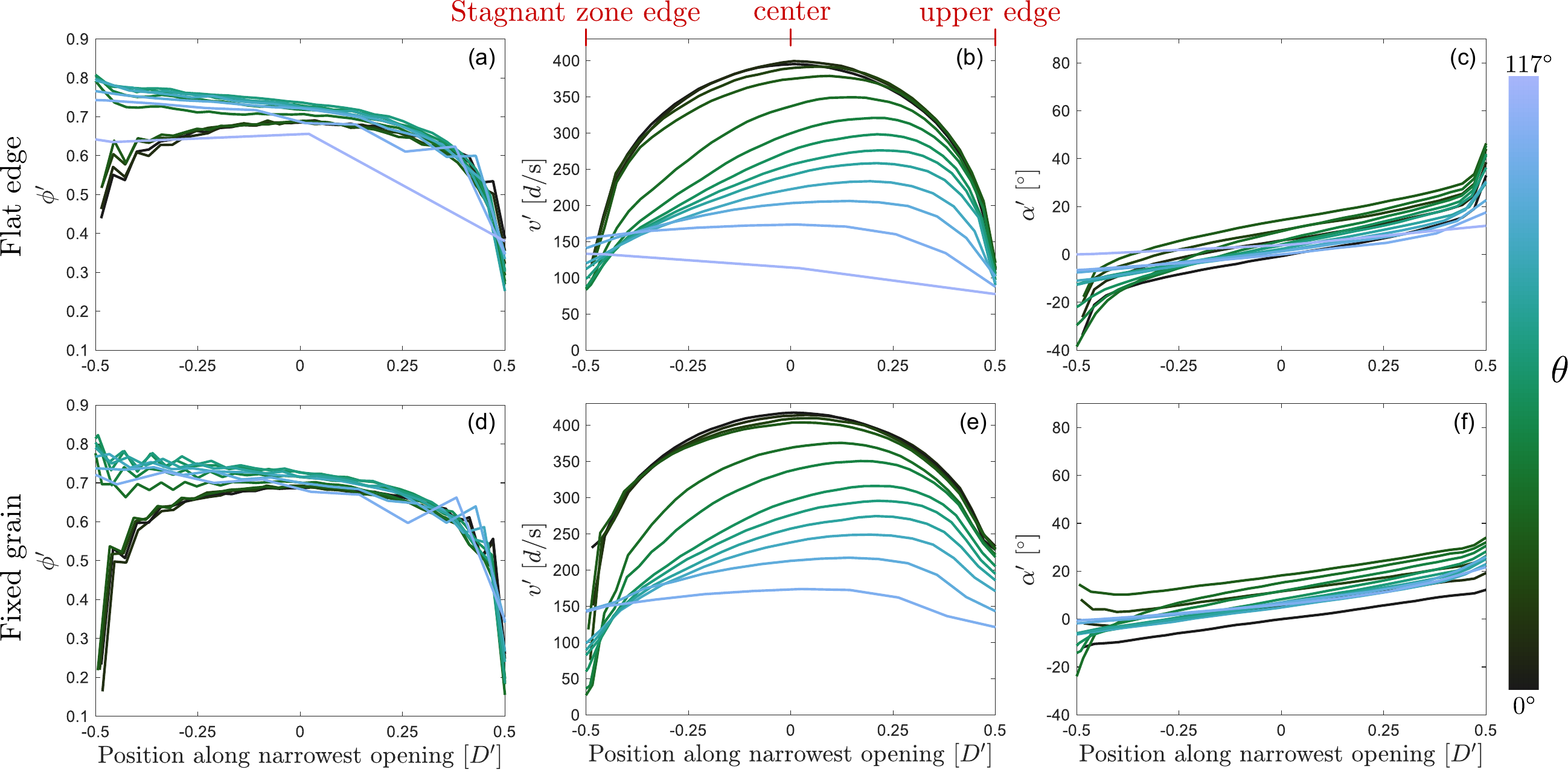}
    \caption{Narrowest opening cross-section profiles of (a,d) packing fraction, (b,e) speed, and (c,f) flow angle for (top row) the flat edge case and (bottom row) the fixed grain case, at varying elevation angles $\theta$ (see color scale).  }
    \label{fig:neworificeprofile}
\end{figure*}

We now analyze the flow along the narrowest opening as demonstrated in Fig.~\ref{fig:neworificemodel}b. (For $\theta<30^{\circ}$, we use the same profiles as analyzed in Sec.~\ref{sec:results} since the orifice \textit{is} the narrowest opening for these angles.) Figure~\ref{fig:neworificeprofile} shows profiles along the narrowest opening of packing fraction $\phi'$, speed $v'$, and flow direction $\alpha'$ with respect to the narrowest opening normal $\hat{n}'$. (See Supplemental Materials for profiles of components $v'_{\perp}$ and $v'_{||}$~\cite{SupplementalMaterials}.) Note that the horizontal axis is normalized to $D'$ since the narrowest opening length varies with $\theta$ according to Eqn.~\ref{eqn:Dprime}. 

The packing fraction and velocity exhibit similar profiles to those with the orifice profile in Fig.~\ref{fig:profiles}. Packing fraction is approximately constant to lowest order with $\theta$, and outpouring speed decreases with $\theta$. Unlike with the conventional orifice analysis, however, $\alpha'$ now collapses with respect to the angle $\theta$ and is close to $0^{\circ}$ (see Fig.~\ref{fig:profiles}c,f).

\subsection{Modeling $\phi'$, $v'$, and $\alpha'$}

\begin{figure*}
    \centering
    \includegraphics[width=1\linewidth]{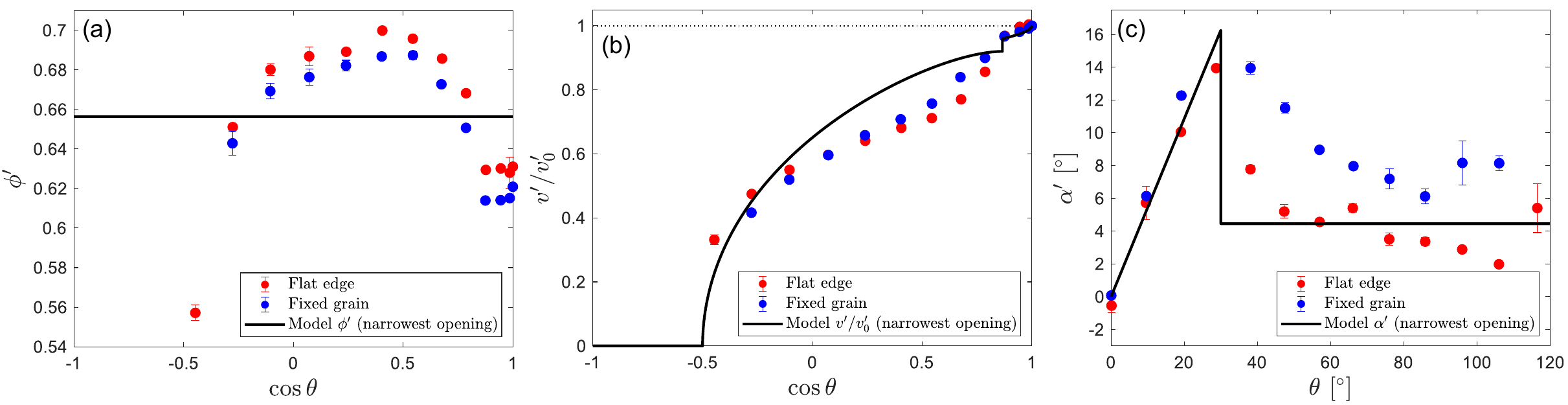}
    \caption{Narrowest opening cross-section: (a) Average packing fraction, (b) average speed normalized to speed at $\theta=0$, and (c) flow angle with respect to cross-section normal for flat edge and fixed grain conditions.
    Black solid lines: model equations for each quantity using $\theta_{\rm S} = 60^{\circ}$ and $\theta_{\rm h} = 120^{\circ}$. }
    \label{fig:newmodelfits}
\end{figure*}

Figure~\ref{fig:newmodelfits} displays averages of the profiles in Fig.~\ref{fig:neworificeprofile} of packing fraction $\phi'$, speed $v'/v'_0$ normalized by the speed at $\theta=0^{\circ}$, and angle $\alpha'$. Our model equations are shown as black solid lines. 
As in Sec.~\ref{sec:model}, we approximate the packing fraction to lowest order as a constant $\phi'(\theta)=0.66$ ($\rho'_0=5.1\;{\rm kg/m}^2$), calculated from the average along the narrowest opening in the flat edge case. For $\alpha'$, we use the model Eqn.~\ref{eqn:modelalphacomplete} for $\theta\leq \pi/2-\theta_{\rm S}$. As a first order approach for $\theta>\pi/2-\theta_{\rm S}$, we take the average of $\alpha'$ for all $\theta>\pi/2-\theta_{\rm S}$ data points in the flat edge case and set $\alpha'={\rm constant}=4.5^{\circ}$. Note that our model for $\alpha'$ \textit{does} include the parameter $\theta_{\rm h}$, but only in range $\theta\leq 30^{\circ}$.

The speed is modeled as proportional to a free-fall lengthscale, as proposed by Zou, et. al. \cite{Zou2020}, but here we use the vertical projection of a segment of length $D'$ (rather than $D$) along the direction of flow at the narrowest opening, which has angle $\alpha'_x$ with the horizontal (see Fig.~\ref{fig:neworificemodel}a). Hence,

\begin{equation}\label{eqn:vvonarrow}
    v'/v'_0 = D'\sin(\alpha'_x),
\end{equation}

 where

\begin{equation}\label{eqn:alphaprimex}
    \alpha'_x = \left\lbrace \begin{array}{ll}
    \pi/2 -\theta +\alpha' , & \theta \leq \pi/2 - \theta_{\rm S}\\
    \theta_{\rm S}+\alpha' , & \theta>\pi/2 - \theta_{\rm S}
    \end{array}\right. .
\end{equation}

In comparison to the average packing fraction along the orifice in Fig.~\ref{fig:profileaverages}a, the packing fraction is higher upstream from the orifice. Correspondingly, as the material is unable to dilate as effectively until it reaches the orifice, the speed measured along the narrowest orifice in Fig.~\ref{fig:newmodelfits}b decays more rapidly with angle in comparison to the speed at the orifice in Fig.~\ref{fig:profileaverages}b. Nevertheless, the free-fall arch consideration of Zou, et. al., is more effective at capturing the speed for the narrowest cross-section instead of the orifice. Lastly, the flow angle is approximately constant and close to $0^{\circ}$ with respect to the narrowest opening normal, implying that the average flow is aligned with the stagnant zone boundary. Each model component, though not a direct match with the data, is an excellent lowest-order fit for the data.

The final result for the normalized flow rate from Eqn.~\ref{eqn:qmodelnarrow} with $\theta_{\rm S}=60^{\circ}$ and $\theta_{\rm h}=120^{\circ}$ is in Fig.~\ref{fig:QQ0} as a thick black line. The narrowest opening model is more successful at capturing the normalized flow rate for the fixed grain and tilted silo available data than the traditional orifice model (thin gray line). This model uses a simple geometric consideration along with measurable properties of the granular material, $\theta_{\rm S}$ and $\theta_{\rm h}$, to describe an effective bottlenecking orifice that constrains the material flow. Moreover, it involves a simpler approach for modeling the outpouring angle and better fits the normalized outpouring speed. We emphasize that prior attempts to model the flow velocity in tilted silo entail empirical or semi-empirical fitting to the data~\cite{Franklin1955,Sheldon2010,Thomas2013,Liu2014b,Kozlowski2023}.

\section{Conclusions}\label{sec:conclusions}

In this work, we presented novel experiments in which the orifice of a vertical silo is rotated by elevating one edge of the orifice. Using high-speed imaging, we analyzed properties of the flow in the vicinity of the orifice to describe the primary factors that contribute to the decrease in flow rate. We demonstrated that the quasi-quiescent stagnant zone near each orifice edge plays an important role for the direction and exit speed of the flowing material. In particular, by constraining a grain at each orifice end, we observe a more clearly defined stagnant zone region and a faster decay of flow rate with orifice angle. Without a fixed grain, the stagnant zones exhibit intermittent creep that contributes grains with horizontal momentum to the flow near the orifice edges. The flow also becomes increasingly horizontal along a slip zone at the bottom edge as tilt angle increases. 

We then developed a novel model based on physics of the flow and measured properties of the granular material. We show that modeling the flow rate based on the flow properties along the orifice cross-section, as has been done with great success in horizontal orifice configurations, is challenging for angled orifice openings. However, the model outcome significantly improves if we consider a new cross-section: the narrowest opening cross-section through which grains pass assuming an idealized stagnant zone. The success of modeling the flow in the fixed grain case and with prior tilted silo results at the narrowest opening implies that the relevant lengthscale for a rotated orifice is $D'$ (Eqn.~\ref{eqn:Dprime}). Correspondingly, one might test the range of validity of this model by modifying properties of the material pertinent for the stagnant zone angle $\theta_{\rm S}$, such as grain-grain friction, grain-boundary friction, size dispersity, shape, and inter-grain cohesion. Moreover, higher-order components of the narrowest-opening model may be applied to more appropriately model the flat edge case by considering the development of the horizontal slip zone at steep angles.

We also showed that the ``flat edge'' and ``fixed grain'' cases are upper and lower bounds, respectively, for the normalized flow rate $Q(\theta)/Q(\theta=0)$ of tilted silos based on prior experimental data. The fixed grain scenario more closely resembles the tilted silo in terms of flow rate. We suggest that the primary distinction between the tilted silo scenario and the flat edge case is that, in the case of the tilted silo, at least one of the boundaries that grains slide along is a stagnant pile of grains, effectively rough and frictional. As we noted in Sec.~\ref{sec:profiles}, the average flow properties of the tilted silo and fixed grain case are similar, suggesting that having at least one effectively rough orifice edge influences the flow similarly. Detailed analysis in this direction is left for future studies.

Our work takes an important step toward understanding the physics of dry granular flow from silos and hoppers with non-conventional opening geometries and boundary conditions. The geometry studied may be of practical use since it allows tilting of the orifice to control flow without tilting the entire silo. 

\section{Declaration of interests}

The authors have nothing to declare.

\section{CRediT (Contributions)}

L.A.P.: Conceptualization, funding acquisition, investigation, validation, visualization, writing -- review and editing. R.K.: Conceptualization, data curation, formal analysis, funding acquisition, investigation, methodology, validation, visualization, writing -- original draft, writing -- review and editing.
\section{Funding sources}

This work was supported by College of the Holy Cross; CONICET (Argentina) [grant PIP-717]; and UNLPam (Argentina) [grant F65]. 

\section{Acknowledgments}

The authors thank C. Manuel Carlevaro for helpful conversations on this work.

\bibliographystyle{elsarticle-num} 

\end{document}


\begin{frontmatter}



\title{Supplemental materials: Flow rate from a vertical silo with a tilted orifice}

\author[label1]{Ryan Kozlowski\corref{cor1}} 
\cortext[cor1]{rkozlows@holycross.edu}

\author[label2]{Luis A. Pugnaloni} 
\affiliation[label1]{organization={Physics Department, College of the Holy Cross},
            addressline={1 College St}, 
            city={Worcester},
            postcode={01610}, 
            state={MA},
            country={United States}}
\affiliation[label2]{organization={Departamento de F\'isica, Facultad de Ciencias Exactas y Naturales,Universidad Nacional de La Pampa, CONICET},
            addressline={Uruguay 151},
            city={Santa Rosa, La Pampa},
            postcode={6300},
            country={Argentina}}                        

\end{frontmatter}




\section{Profiles of $v_{\perp}$ and $v_{||}$ along orifice and narrowest opening cross-sections }

\begin{figure}[b]
    \centering
    \includegraphics[width=0.7\linewidth]{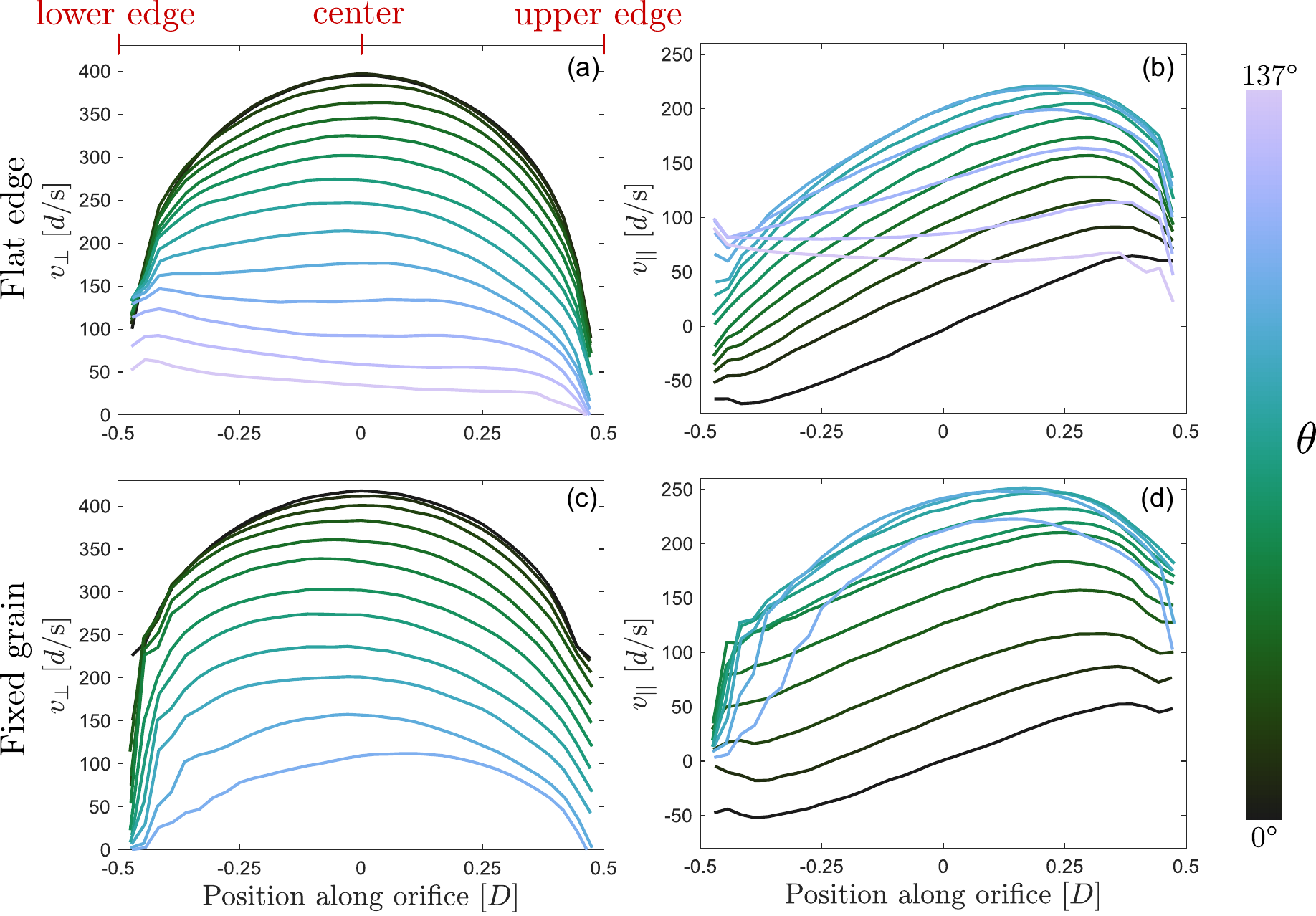}
    \caption{Profiles along the orifice of (a,c) outpouring velocity $v_{\perp}$ and (b,d) the component of velocity along the orifice $v_{||}$ for (top row) the flat edge case and (bottom row) the fixed grain case, at varying elevation angles $\theta$ (see color scale).}
    \label{fig:profilevperpvpar1}
\end{figure}

\begin{figure}
    \centering
    \includegraphics[width=0.7\linewidth]{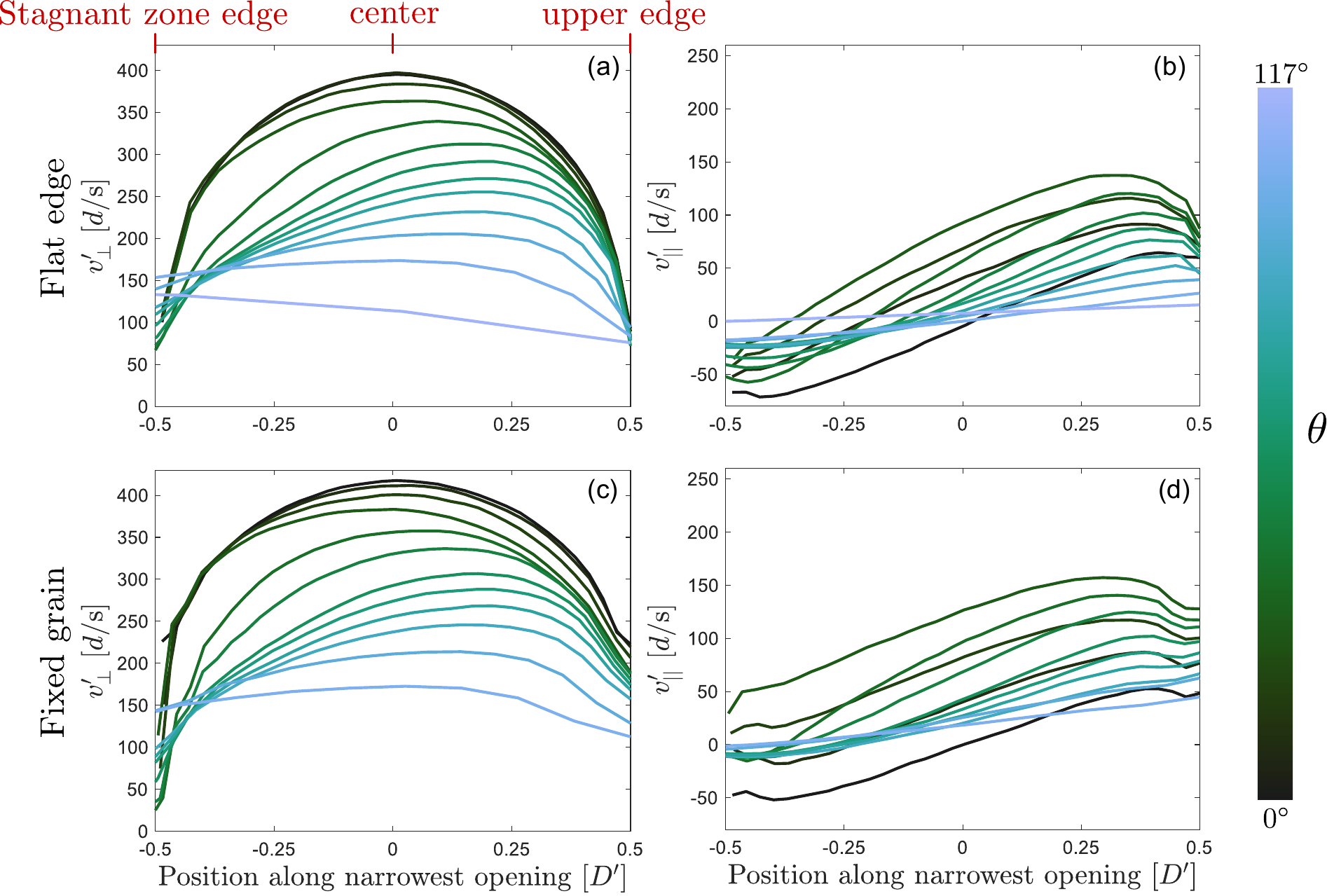}
    \caption{Profiles along the narrowest opening cross-section of (a,c) the component of velocity $v'_{\perp}$ directed along the cross-section normal and (b,d) the component of velocity along the cross-section $v'_{||}$ for (top row) the flat edge case and (bottom row) the fixed grain case, at varying elevation angles $\theta$ (see color scale).}
    \label{fig:profilevperpvpar2}
\end{figure}

Profiles of each component of velocity $v_{\perp}$ and $v_{||}$ with respect to the cross-section normal vector are shown in Fig.~\ref{fig:profilevperpvpar1} (cross-section along the orifice) and Fig.~\ref{fig:profilevperpvpar2} (narrowest opening cross-section). $v_{\perp}$ is the component perpendicular to the cross-section, or aligned with the normal vector of the cross-section. $v_{||}$ is the component parallel to the cross-section, or aligned with the cross-section itself. Refer to Secs. 2 and 3 in the main text for the definitions of $v_{\perp}$ and $v_{||}$ and the protocol for measuring profiles, as well as Figs. 6 and 15 for profiles of packing fraction, speed, and flow angle.

\section{Heat maps of average flow}

\begin{figure}
    \centering
    \includegraphics[width=1\linewidth]{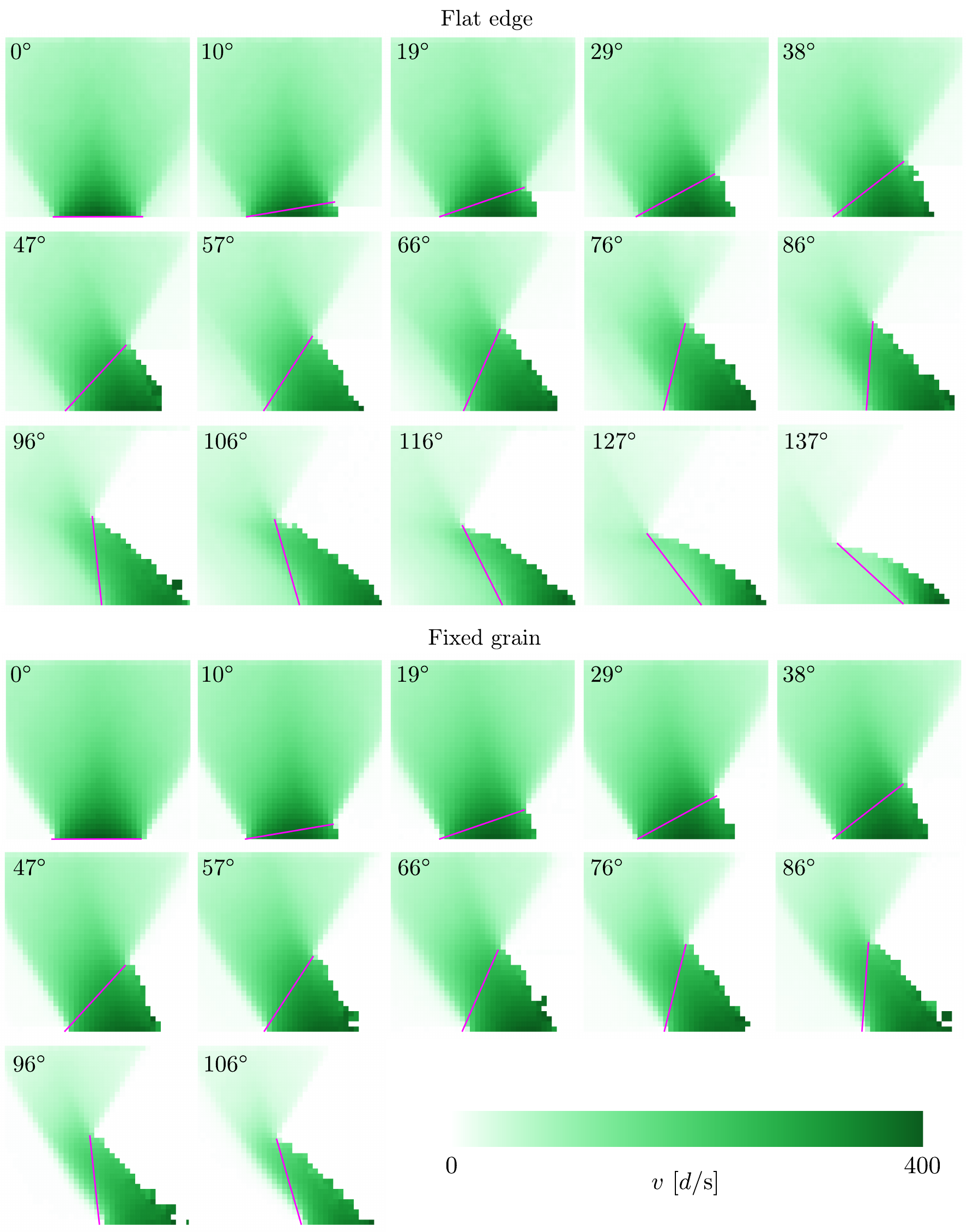}
    \caption{Heat map of speed $v$ for flat edge and fixed grain cases. Labels indicate the orifice angle $\theta$.}
    \label{fig:heatmapspeed}
\end{figure}

\begin{figure}
    \centering
    \includegraphics[width=1\linewidth]{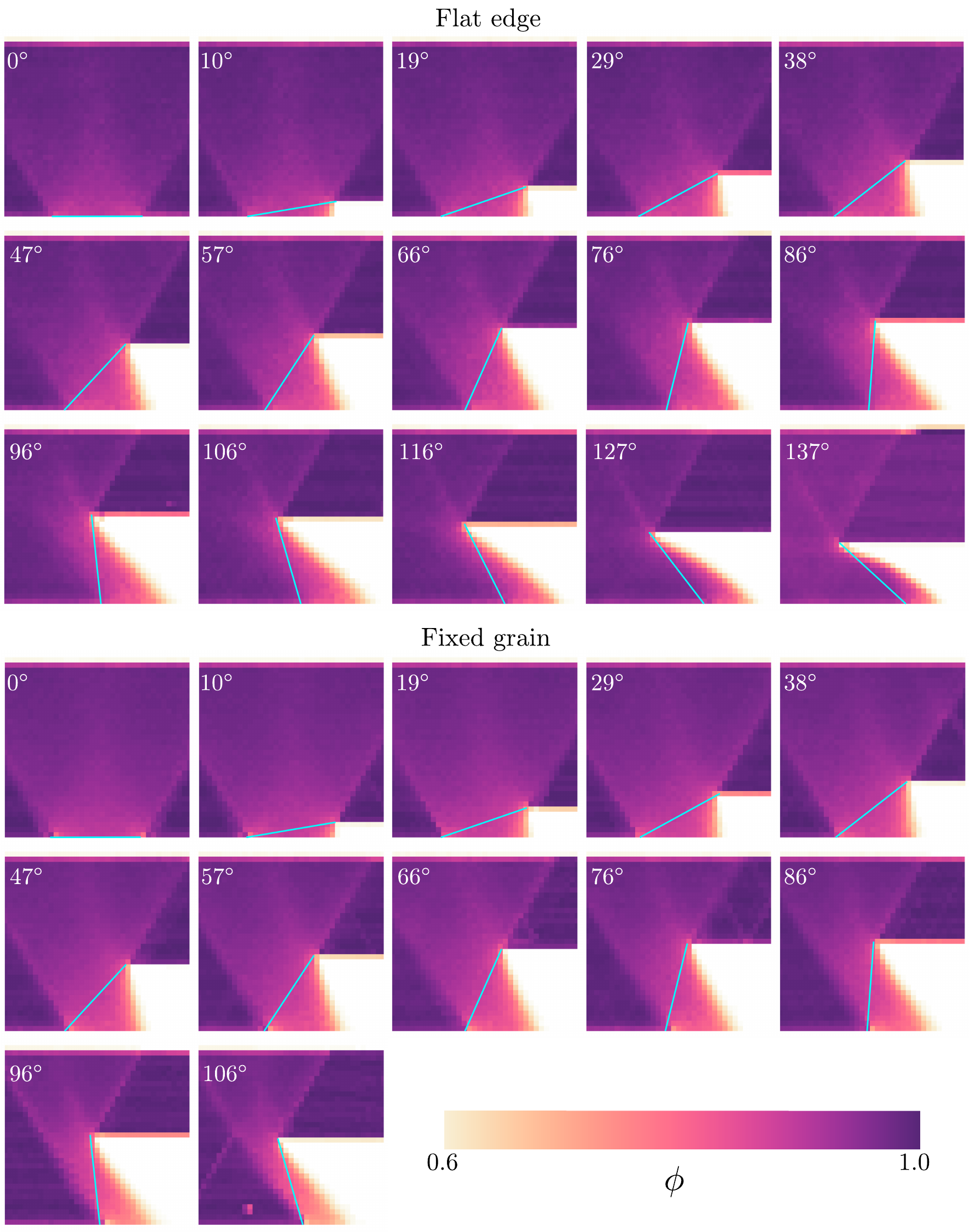}
    \caption{Heat map of packing fraction $\phi$ for flat edge and fixed grain cases.  Labels indicate the orifice angle $\theta$.}
    \label{fig:heatmappf}
\end{figure}

The average speed $v$ and $\phi$ heat maps for all $\theta$, in both the flat edge and fixed grain cases, are shown in Fig.~\ref{fig:heatmapspeed} and~\ref{fig:heatmappf}. The heat maps are obtained by segmenting experimental images into $1d\times1d$ bins and averaging the packing fraction $\phi$ and speed $v$ (weighting the speed average based on contribution to packing fraction in each bin) of grains overlapping each bin throughout all frames. 

\section{Supplemental videos}

The following sample videos from experiment trials in this manuscript are available as Supplemental Materials. The imaging frame rate for each video is given below. All videos are played back 25$\times$ slower than real time.

\begin{itemize}
    \item \textbf{flat edge 0 degrees}: 2000 frames per second.
    \item \textbf{flat edge 10 degrees}: 2000 frames per second.
    \item \textbf{flat edge 19 degrees}: 2000 frames per second.
    \item \textbf{flat edge 29 degrees}: 2000 frames per second.
    \item \textbf{flat edge 38 degrees}: 2000 frames per second.
    \item \textbf{flat edge 47 degrees}: 2000 frames per second.
    \item \textbf{flat edge 57 degrees}: 2000 frames per second.
    \item \textbf{flat edge 66 degrees}: 2000 frames per second.
    \item \textbf{flat edge 76 degrees}: 2000 frames per second.
    \item \textbf{flat edge 86 degrees}: 1000 frames per second.
    \item \textbf{flat edge 96 degrees}: 1000 frames per second.
    \item \textbf{flat edge 106 degrees}: 1000 frames per second.
    \item \textbf{flat edge 116 degrees}: 750 frames per second.
    \item \textbf{flat edge 127 degrees}: 750 frames per second.
    \item \textbf{flat edge 137 degrees}: 750 frames per second.
    \item \textbf{fixed grain 0 degrees}: 2000 frames per second.
    \item \textbf{fixed grain 10 degrees}: 2000 frames per second.
    \item \textbf{fixed grain 19 degrees}: 2000 frames per second.
    \item \textbf{fixed grain 29 degrees}: 2000 frames per second.
    \item \textbf{fixed grain 38 degrees}: 2000 frames per second.
    \item \textbf{fixed grain 47 degrees}: 2000 frames per second.
    \item \textbf{fixed grain 57 degrees}: 2000 frames per second.
    \item \textbf{fixed grain 66 degrees}: 2000 frames per second.
    \item \textbf{fixed grain 76 degrees}: 2000 frames per second.
    \item \textbf{fixed grain 86 degrees}: 1000 frames per second.
    \item \textbf{fixed grain 96 degrees}: 1000 frames per second.
    \item \textbf{fixed grain 106 degrees}: 1000 frames per second.
    \item \textbf{fixed grain 116 degrees}: 750 frames per second.
\end{itemize}
